%% file: manuscript.tex
\documentclass{article}
\usepackage[T1]{fontenc}
\usepackage{graphicx,booktabs,calc,array,longtable}
\usepackage{hyperref}
\usepackage[preprint]{icml2026}
\usepackage{amsmath,amssymb}
\hypersetup{hidelinks,pdftitle={Common-DAG Support from Targeted Covariance Changes under Confounding},pdfsubject={Preprint},pdfauthor={Peida Li}}
\providecommand{\tightlist}{\setlength{\itemsep}{0pt}\setlength{\parskip}{0pt}}

\icmltitlerunning{Common-DAG Support under Confounding}
\renewcommand{\today}{November 11, 2026}
\begin{document}
\twocolumn[
\icmltitle{Common-DAG Support from Targeted Covariance Changes under Confounding}
\begin{icmlauthorlist}
\icmlauthor{Peida Li}{pku}
\end{icmlauthorlist}
\icmlaffiliation{pku}{School of Basic Medical Sciences, Peking University}
\icmlcorrespondingauthor{Peida Li}{2410301207@stu.pku.edu.cn}
\vskip 0.3in
]
\printAffiliationsAndNotice{}
\begin{abstract}
\input{abstract.tex}
\end{abstract}
\input{main.tex}
\clearpage
\typeout{MAIN_PAGE_COUNT=\number\numexpr\value{page}-1\relax}
\section*{Impact Statement}
This work develops statistical methods for directed relations in confounded systems. Interpreting their graph outputs requires matching the intervention model and parent information to the application. The empirical analyses use public experimental resources.

\input{references.tex}
\clearpage
\onecolumn
\appendix
\raggedbottom
\setcounter{figure}{0}
\renewcommand{\thefigure}{S\arabic{figure}}
\renewcommand{\theHfigure}{S\arabic{figure}}
\input{appendices.tex}
\end{document}

%% file: abstract.tex
Targeted covariance changes couple structural rows through a common directed acyclic graph. We study the graph support determined by these experiments when structural errors may be correlated. A positive-gap four-node family has a shortcut identified by common acyclicity. A response indegree of at most two additionally gives unique complete support and two compatible coefficient matrices. The same experiments admit a stable feedback model with the shortcut absent. The result extends to connected graphs with repeated branching and merging. Finite response-column alternatives yield exact bounded-parent compatibility decisions through Horn propagation and partial-target completion. Certified ancestry then supplies an observed-ancestor covariance restriction. Gaussian regression and Fieller inversion reduce inference to a scalar confidence set, and quadratic elimination retains the shared nuisance coefficients with finite-sample edge-error control. In 2048 Gaussian conditions, the scalar procedure reports 127/128 positive edges at gap 1/4 and 524288 observations per cell. Its decisions agree with the joint outer-polynomial reference on the same inputs in all 2022 qualified conditions, with zero reports in 512 null conditions. Exact 16-, 32- and 64-node evaluations verify support and target-coverage conclusions. A root-target characterization connects the identifying information to intervention design.

%% file: main.tex
\section{Introduction}\label{introduction}

Known intervention targets constrain structural-error covariance changes. With correlated errors, a contrast can retain several possible response columns, each consistent with its target. These columns must describe one acyclic structural system. A locally invariant residual fit can require column choices whose joint support contains a cycle. \citep{ref1,ref2,ref3,ref4,ref5}

We ask which direct edges are necessary in every common DAG compatible with the observed covariance experiments. Our model allows a target to change both its error variance and its correlations with other errors, while preserving the off-target error block. Each informative contrast gives at most two normalized response columns. Selecting these columns jointly exposes information carried by common acyclicity.

A four-node family makes this information explicit. Three targeted contrasts identify the chain u-to-x-to-v-to-w and its positive-gap shortcut u-to-w under a two-parent response condition. Two compatible shortcut coefficients have opposite signs. A stable cyclic representation and a locally invariant missing-edge residual remain compatible with the same experiments. The source of identification is the requirement that the response columns belong to one DAG.

We develop three consequences. First, continuous core and connected-graph families establish support identification with coefficient ambiguity and a matching fixed-dimensional sample order. A root-target characterization gives an exact design boundary for this compatibility class. Second, finite response compatibility and partial-target completion give exact bounded-parent decisions. Third, certified ancestry supplies a Gaussian confidence-inversion procedure whose scalar elimination preserves shared nuisance coefficients.

The experiments follow these consequences. Connected population evaluations examine support, target coverage and parent domains. Shared-input Gaussian comparisons measure the power obtained from ancestry and confidence-region geometry, including agreement with a joint polynomial formulation. Public measured acquisitions illustrate the intervention-design requirements and support the application-specific response analyses in Appendix D.

\section{Model and related work}\label{model-and-related-work}

Let \(X=BX+\epsilon\), where \(B\) has zero diagonal and acyclic support. Write \(W=I-B\) and \(A=W^{-1}\). Structural errors have an arbitrary positive-definite covariance. The control covariance is \(\Sigma_0\). A known single-target experiment j has covariance \(\Sigma_j\) and satisfies

\[
[W(\Sigma_j-\Sigma_0)W^\top]_{-j,-j}=0. \tag{1}
\]

The targeted error can change its variance and its correlations with other errors. All experimental covariances share the same B. The Gaussian sampling interface uses independent rows within each cell and separately estimated cell means. Reusing a control covariance across contrasts is permitted.

For a query u-to-w, let K be a qualified complete parent domain for w, containing u, and put P=K excluding u. The missing-edge class imposes pa(w) contained in P. A local response-indegree bound can instead define a union over candidate P. Optional qualified roots exclude w. Other structural rows remain unrestricted. Write \(\mathcal E\) for the covariances, target labels and declared structural restrictions, and \(\mathcal M_{\rm DAG}(\mathcal E)\) for their compatible DAG matrices. For a nonempty class, the inferential target is

\[
E_\cap(\mathcal E)=\bigcap_{B\in\mathcal M_{\rm DAG}(\mathcal E)}\operatorname{supp}(B).
\]

An edge e belongs to \(E_\cap\) exactly when its missing-edge class \(\mathcal M_{0,e}\) is empty.

Invariant causal prediction combines invariant residual laws with predictor--error independence. WVM tests a residual-distribution objective and has a finite-sample uniform deviation theorem under its function-class and tail assumptions. Our identifying family gives this objective a zero missing-edge value while common DAG compatibility forces the shortcut. \citep{ref1,ref2}

Causal Dantzig derives inner-product invariance under additive interventions. Our ancestor qualification selects the observed cross-moments preserved by the off-target error block in (1). MuDo-CoM instead assumes a common dependent-component covariance transformed by domain-specific diagonal scaling. Appendix L gives the formula-level comparisons. \citep{ref3,ref4}

LinADMG identifies structural coefficients after recovering a noise representation up to permutation and scaling. Its Lemma 4.1 recovers B under those representation assumptions. Our two compatible normalized responses differ by an explicit non-monomial shear, while their DAG support remains uniquely identified. Appendix L gives the matrices and identity. \citep{ref21}

LLC and inverse sparse regression supply the experimental-response inversion lineage. Fieller inversion and Horn propagation are inherited mathematical tools. Changes-based and known-target methods describe experimental graph information under their intervention semantics. Our decision combines finite response alternatives, exact structural-row equations and common acyclicity in (1). \citep{ref5,ref6,ref7,ref8,ref22,ref23,ref24}

\par

\noindent

\begin{table*}[t]
\textbf{Table 1. Claims, assumptions and evidence.}

\begin{center}
\small
\begin{tabular}{@{}p{\dimexpr0.260000\linewidth-2\tabcolsep\relax}p{\dimexpr0.380000\linewidth-2\tabcolsep\relax}p{\dimexpr0.360000\linewidth-2\tabcolsep\relax}@{}}
\toprule
\textbf{Claim} & \textbf{Model or information} & \textbf{Evidence} \\
\midrule
Finite target-column candidates & Equation (1), informative contrast, nonzero diagonal anchor & Proposition 1 \\
Exact missing-edge decision & Finite parent-column candidates, valid parent domain and root equations & Theorem 1 and partial-target completion \\
Root-target identification boundary & Known roots, root targets, arbitrary coefficients and correlated errors & Proposition 2, 24 rational systems \\
Necessary shortcut & Positive-gap four-node family, unrestricted response indegree & Corollary 1, compatible missing-edge feedback model \\
Unique complete support & The same family, response indegree at most two & Theorem 2 and its candidate enumeration \\
Simultaneous positive-edge control & Gaussian cells, valid qualification and covering regions & Theorems 4--5, 512 null conditions \\
Profile confidence computation & Certified order, Fieller set and shared variance constraints & 2048 paired inputs, single-box ablation and joint outer-system checks \\
Connected support and target coverage & Core targets, singleton downstream targets and response indegree two & Theorem 3, 96 population conditions \\
\bottomrule
\end{tabular}
\end{center}
\end{table*}
\par\medskip

\section{Response factors and common compatibility}\label{response-factors-and-common-compatibility}

\subsection{Two normalized candidates from a target contrast}\label{two-normalized-candidates-from-a-target-contrast}

Put \(S_j=\Sigma_j-\Sigma_0\) and \(a_j=Ae_j\). Equation (1) is equivalent to a factorization

\[
S_j=a_jh_j^\top+h_ja_j^\top,\qquad (a_j)_j=1. \tag{2}
\]

Let \(c=(S_j)_{jj}\), \(s=S_je_j\) and \(D=ss^\top-cS_j\). For c nonzero, equation (2) gives \(D=dd^\top\) with \(d_j=0\), where \(d=h_j-(c/2)a_j\). Hence

\[
\mathcal C_j=\{(s+d)/c,(s-d)/c\}. \tag{3}
\]

A rank-zero D gives a singleton. A realizable rank-one positive-semidefinite D determines d up to sign. Each candidate is checked against the full contrast. Several contrasts for one target intersect their candidate sets. A finite-sample anchor interval containing zero retains the universal anchored column region.

\textbf{Proposition 1. Complete targeted factor candidates.} Under (1), every informative nonzero-anchor target response belongs to (3). Conversely, every normalized candidate in (3) has a partner h satisfying (2).

For the converse, set h=s-(c/2)a and expand \(ah^\top+ha^\top\) using \(D=dd^\top\). This recovers \(S_{j}\). The construction determines the complete column candidate set from the observed population contrast.

\subsection{Acyclic selection and Horn propagation}\label{acyclic-selection-and-horn-propagation}

For a selected response column \(a_{j}\), draw j-to-i when i differs from j and \((a_j)_i\) is nonzero. A complete selection is compatible with an acyclic structural model exactly when this response-support graph is acyclic. Appendix K gives both directions of this compatibility equivalence.

This selection problem has a sink-removal solution. A node is eligible when one of its column candidates points only to nodes already removed. Select that candidate and remove the node. Every feasible selection has an eligible sink. Removing any eligible node preserves feasibility of the remaining independent column choices. Thus the procedure succeeds exactly when a compatible acyclic selection exists.

Each candidate support defines a Horn implication from its outgoing neighbors to its owner. Maintaining unsatisfied-neighbor counts implements the classical propagation principle. With input support size H, a compatibility decision costs \(O(p+H)\). Dense two-candidate columns give \(H=O(p^2)\). \citep{ref8}

\subsection{Partial targeting and bounded-parent queries}\label{partial-targeting-and-bounded-parent-queries}

A missing u-to-w representation obeys

\[
A_{wj}-\sum_{r\in P}\beta_rA_{rj}=\mathbf1\{j=w\}. \tag{4}
\]

When P consists of known target columns, enumerate its at most \(2^{|P|}\) candidate selections. In an acyclic choice, \(A_{PP}\) is unit triangular after permutation and is invertible. These columns determine

\[
\beta^\top=A_{wP}(A_{PP})^{-1}. \tag{5}
\]

Filter every available column by (4) and any declared root equations, then apply Horn compatibility. Every actual missing representation supplies one enumerated branch. Every successful branch supplies an explicit compatible missing representation.

Untargeted columns admit a canonical null completion. Keep the known columns. For an unknown column j in P use \(e_j+\beta_je_w\). For every other unknown column use \(e_{j}\). The response-row equation remains valid. Each added off-diagonal entry is an original structural parent edge in row w, so a topological order of the original missing representation also orders the completed matrix. The observed target contrasts retain their selected columns and their legal transformed blocks.

\textbf{Theorem 1. Partial-target missing-edge decision.} Suppose the possible remaining parents P have finite known response-column candidates. With the declared root and response-parent restrictions, a compatible missing-edge DAG exists exactly when some branch of (5), completed and filtered as above, passes Horn propagation. With at most two candidates per known column, a query costs

\[
O\{2^{|P|}(p^2+|P|p+|P|^3)\}. \tag{6}
\]

A valid local indegree bound d permits enumeration of every subset P of the other p-2 coordinates with size at most d.~The number of coefficient branches is at most \(\sum_{k=0}^d\binom{p-2}{k}2^k\). The theorem applies directly when those parent columns are observed. For a more general fixed spanning-target set Z, its candidate block \(A_{PZ}\) must enclose every invertible branch required by the true missing model.

\subsection{Information in root-target designs}\label{information-in-root-target-designs}

Partition the coordinates into known structural roots R and the remainder N, with targets \(J\subseteq R\). Each target has a nonempty candidate set satisfying all its contrasts and root equations. Write \(C:=A_{NR}\), so \(A_{:R}=\binom{I_R}{C}\). Complete the unobserved columns of this root-to-N response block freely. Allow every acyclic coefficient matrix on N, arbitrary real coefficients and arbitrary positive-definite error covariances. Coordinates in N may have empty parent sets.

\textbf{Proposition 2. Root-target compatibility.} Let \(u\in J\), \(w\in N\), with every other coordinate allowed as a parent of w. For fixed root-to-N responses, a missing-edge DAG exists exactly when \(C_{wu}=0\), or some \(m\in N\setminus\{w\}\) has \(C_{mu}\ne0\). Across candidate choices, u-to-w is necessary precisely when every u column has a nonzero w entry and zero entries at every other N coordinate. For \(u\in R\setminus J\), completing \(C_{:u}=0\) gives the full response column \(A_{:u}=e_u\) and a missing-edge DAG.

For an acyclic matrix T on N, set \(B_{NR}=(I-T)C\) and \(B_{NN}=T\). This preserves every root response. Taking only \(T_{wm}=C_{wu}/C_{mu}\) nonzero removes the queried edge. The converse follows from \(C_{wu}=\sum_{m\in N\setminus\{w\}}B_{wm}C_{mu}\). Appendix K.6 proves preservation of all covariances and relates the construction to response-preserving ambiguity in LLC. \citep{ref22}

The zero-response requirements are population equalities. Their sampling uncertainty and the unbounded mediator coefficient govern statistical separation. Section 4's internal targets supply the additional response-column information used by its identifying family.

\section{Unique complete support and a stable missing-edge feedback model}\label{unique-complete-support-and-a-stable-missing-edge-feedback-model}

Consider the chain u-to-x-to-v-to-w with a possible shortcut u-to-w. Each target contrast admits two response columns. Choosing the alternate v column points back to x and closes an x-v response cycle. Choosing alternate x while keeping v physical gives w three parents. Acyclicity and the two-parent condition therefore fix the x and v columns, while either u column remains possible. These two surviving choices have the same positive-gap support and different shortcut coefficients. Local residual invariance can retain a missing-edge fit because it tests the response row separately from these common response-column choices.

Order the physical coordinates as (u,x,v,w). Take a,c,b,k,l positive, e negative, k\textgreater c, kl\textless1 and s nonnegative. Set

\[
\begin{aligned}
a_u&=(1,a,ac,bac+s)^\top,& v_u&=(1,a,ac,eac)^\top,\\
a_x&=(0,1,c,bc)^\top,&v_x&=(0,1,k,ek)^\top,\\
a_v&=(0,0,1,b)^\top,&v_v&=(0,l,1,e)^\top.
\end{aligned} \tag{7}
\]

The physical DAG has coefficients \(B_{xu}=a\), \(B_{vx}=c\), \(B_{wv}=b\) and \(B_{wu}=s\). Node w is untargeted. For j in \{u,x,v\}, use

\[
\begin{aligned}S_j&=a_jv_j^\top+v_ja_j^\top,\\
q_j&=\frac{3}{4\max_i\sum_h|(S_j)_{ih}|},\quad
\Sigma_j=I+q_jS_j.\end{aligned}\tag{8}
\]

Use centered Gaussian laws for these covariances. Every covariance has eigenvalues between 1/4 and 7/4. Under the physical W, each transformed contrast is supported on its targeted error row and column.

\textbf{Theorem 2. Complete-support identification beyond local row laws.} In the covariance family (7)--(8), assume w has at most two structural parents. Every compatible DAG has the chain coefficients (a,c,b) and shortcut coefficient

\[
B_{wu}\in\{s,(e-b)ac\}. \tag{9}
\]

For s\textgreater0 its complete support is uniquely the chain plus u-to-w. At s=0 the chain is the unique minimum-cardinality compatible support. For every s, a stable zero-diagonal cyclic representation has \(B_{wu}=0\) and generates the same covariance experiments. The missing row also preserves its Gaussian residual marginal across all cells and its legal joint law with U across the non-u targets.

Appendix K gives the complete candidate enumeration and the explicit stable feedback construction.

\textbf{Corollary 1. Queried-edge identification from acyclicity.} With unrestricted response indegree, the compatible shortcut coefficients in the same family are s, (e-b)ac, s+(b-e)ak and (e-b)a(c-k). All are nonzero for s\textgreater0. Thus u-to-w belongs to \(E_\cap\). Theorem 2's degree condition additionally identifies complete support. Appendix K.3 gives the enumeration and positive-definite baseline extension.

\textbf{Proposition 3. Fixed strong edges and sample difficulty.} Fix p=4, a compact parameter class with positive margins for a,c,b,l,\textbar e\textbar,k-c and 1-kl, and error levels \(\alpha\),\(\beta\) with \(\alpha\)+\(\beta\)\textless1. There are positive constants C,c independent of \(s\in(0,s_{\max}]\) such that the following holds. If the retained arithmetic enclosure error is at most c s and

\[
 n\ge C s^{-2}\log\frac{mp^2}{\alpha\wedge\beta},
\]

the conservative common-DAG query reports u-to-w with probability at least 1-\(\beta\), while its simultaneous type-I error is controlled by the declared \(\alpha\) allocation. The compatible strong representative has four edge coefficients bounded away from zero independently of s. Any test distinguishing its positive-s experiment from the missing physical chain at zero with these fixed nontrivial errors requires n of order at least \(s^{-2}\).

Appendix C gives the continuous margins and KL reduction. Appendix K records the uniform candidate and parent-inverse argument.

\subsection{Connected complete-support families}\label{connected-complete-support-families}

Attach a downstream DAG with coefficient matrix D to w through coefficients c.~The core is ancestor-closed, \(\|D\|_2\le\rho<1\) and \(\|c\|_2\le(1-\rho)/16\). Lift the three core contrasts through its response and target w and every added coordinate with an informative rank-one contrast. These added target columns have singleton candidates. Appendix E gives the block matrices, uniform covariance bounds and complete proof.

\textbf{Theorem 3. Connected complete support.} Suppose all added target columns are available and w has at most two structural parents. For s\textgreater0 every compatible DAG has the complete support of the physical connected graph. Exactly two coefficient matrices are compatible, with fixed c and D and shortcut coefficients s and (e-b)ac. The same experiments admit a stable zero-diagonal feedback representation with the shortcut absent.

Removing any of the three core targets permits an explicit acyclic missing-shortcut model. The construction retains three ambiguous core columns as the observed dimension grows. Its sampled core marginal is the qualified experiment in Section 6.1.

\section{Simultaneous finite-sample certificates}\label{simultaneous-finite-sample-certificates}

\subsection{Covariance and response regions}\label{covariance-and-response-regions}

There are m covariance cells of dimension p.~Coordinate forms and all pairwise sums and differences give \(p^{2}\) one-dimensional forms per cell. For a form with unbiased sample variance v and \(\nu=n-1\) degrees, \(\nu v/\sigma^2\) has a chi-square law. Choose the least integer t with \(2^t\alpha_\Sigma\ge2mp^2\). Laurent--Massart concentration and \(e^t\ge2^t\) give simultaneous tail probability at most \(\alpha_\Sigma\). \citep{ref9}

Let \(r_U\) enclose \(\sqrt{\nu t}\) from above, \(a_L=\nu-2r_U\) and \(b_U=\nu+2r_U+2t\). If the actual sample variance is enclosed by \([v_L,v_U]\), use

\[
\sigma^2\in\left[\frac{\nu\max(0,v_L)}{b_U},
\frac{\nu\max(0,v_U)}{a_L}\right] \tag{10}
\]

when \(a_L>0\), retaining an infinite upper endpoint otherwise. Polarization converts the covered form variances to entrywise covariance boxes. Each shared control is covered on the same event. Several nested budgets can be included as distinct cell-budget pairs in this allocation.

Outward arithmetic propagates the contrast boxes through \(D=ss^\top-cS\). A positive pivot \(D_{kk}\) gives intervals for \(d_k=\sqrt{D_{kk}}\) and \(d_i=D_{ik}/d_k\). Equation (3) supplies two response boxes. An uncertain pivot uses the coordinate bounds \(|d_i|\le\sqrt{\max(0,D_{ii}^{\rm upper})}\). An anchor containing zero retains the universal anchored column region. These operations preserve every population candidate on the simultaneous covariance event.

\subsection{Ancestry contraction and interval queries}\label{ancestry-contraction-and-interval-queries}

Mandatory response supports certify shared ancestor relations. A common topological order forces every coordinate pointing to an ancestor to zero, tightening each outward box. For every parent branch, enclose the coefficients in (5), remove columns only when their row equation excludes zero, and run Horn propagation on mandatory supports. Singular inverses and unfinished branches remain unresolved. Appendix K gives the branch-preservation argument.

\textbf{Theorem 4. Simultaneous positive-edge control.} Under the Gaussian model and correct structural qualification, every reported edge is present in the true DAG on the common covariance event. The probability of at least one false report is at most \(\alpha_\Sigma\). A separate qualification failure event with budget \(\alpha_A\) gives total bound \(\alpha_\Sigma\)+\(\alpha_A\).

At a true missing edge, its true parent-column choices occur in the branch list and its coefficient belongs to the outward inverse enclosure. Its actual columns remain after ancestry contraction and row filtering. Their mandatory supports form a subgraph of an acyclic response graph. Horn propagation therefore succeeds for that branch. All-branch exclusion is impossible on the covering event, proving the theorem simultaneously across queries using those regions.

\textbf{Algorithm 1. Common-DAG missing-edge certification.}

\begin{enumerate}
\def\labelenumi{\arabic{enumi}.}
\tightlist
\item
  Receive cell observations, known targets, a qualified response-parent class and the error allocation.
\item
  Enclose actual sample covariances and construct simultaneous population covariance boxes.
\item
  Construct outer response-column regions, retaining uncertain anchors and model-qualification states.
\item
  Propagate common mandatory ancestor relations and intersect forced backwards coordinates with zero.
\item
  Enumerate the declared missing-edge parent and spanning-column branches.
\item
  Contract each branch, enclose its parent coefficients, filter necessary row equations and run mandatory-support Horn propagation.
\item
  Return an edge certificate on complete branch exclusion. Return an unresolved query for every retained outer branch.
\end{enumerate}

\subsection{Observed-ancestor covariance and scalar confidence inversion}\label{observed-ancestor-covariance-and-scalar-confidence-inversion}

Consider a qualified four-coordinate query with order \(u<x<v<w\) and complete parent domain \{u,x,v\} for w. The four cells 0,u,x,v are sampled independently, with independent rows within each. The same control sample is reused across comparisons. A missing u-to-w edge has residual \(R=Y-tX-bV=\epsilon_w\). Node X depends only on errors upstream of v. The unchanged error block across control and target v therefore gives common covariance for (X,R). Adding tX yields

\[
\operatorname{Cov}_0\begin{pmatrix}X\\Y-bV\end{pmatrix}
=\operatorname{Cov}_v\begin{pmatrix}X\\Y-bV\end{pmatrix}. \tag{11}
\]

Let \(\mathcal F_{\rm anc}\) contain coefficients satisfying (11) and the population residual-variance equalities below. Then \(\pi_{t,b}(\mathcal M_{0,u\to w})\subseteq\mathcal F_{\rm anc}\) in this qualified class. Algorithm 2 encloses these necessary conditions. Exclusion certifies the edge, while a retained pair leaves the query unresolved. A full DAG witness additionally requires response-column compatibility.

Equation (11) eliminates t. In the identifying family its cross-covariance contrast is \(q_vl(b_0-b)\), identifying \(b_0\). This is the ancestor information added to the root-based residual comparison.

Fit separate intercepts and slopes in cells 0 and v. Let \(H_g\) be the centered X scatter, \(d_{Y}\) and \(d_{V}\) the corresponding between-cell differences in regression slopes for Y and V, and \(Q_{g}\)(b) the residual sum of squares for regressing Y-bV on X in cell g. With \(\nu_F=n_0+n_v-4\), the true missing-edge coefficient satisfies

\[
F(b)=\frac{(d_Y-bd_V)^2}
{(H_0^{-1}+H_v^{-1})\{Q_0(b)+Q_v(b)\}/\nu_F}
\sim F_{1,\nu_F}. \tag{12}
\]

Conditional on the two X designs, the common Gaussian slope and residual variance give independent numerator and pooled residual sum of squares. Classical Fieller-type inversion of \(F(b)\le c_F\) gives a scalar quadratic confidence set \(C_{b}\). \citep{ref24} Bounded intervals, disconnected rays, the whole line and empty sets retain their ordinary set meanings. All components are covered by the subsequent calculation.

For g in \{u,x\}, let \(h_{g}\)(t,b) be the sample residual variance of Y-tX-bV. A true missing row has equal variance in cells 0,u,x. Two variance-ratio tests give four constraints

\[
h_g(t,b)-c_gh_0(t,b)\le0,\qquad
h_0(t,b)-c_gh_g(t,b)\le0, \tag{13}
\]

Here \(h_{g}\) is the unbiased within-cell variance of Y-tX-bV. Appendix F gives its polynomial coefficients.

For a bounded b interval, outward coefficient bounds produce an outer quadratic sublevel set on each t half-line. For a coefficient interval \([A_L,A_U]t^2+[B_L,B_U]t+[C_L,C_U]\), the lower polynomial uses \((A_L,B_L,C_L)\) on \(t\ge0\) and \((A_L,B_U,C_L)\) on \(t\le0\). Intersect the four full-line outer sets. Empty intersection excludes the entire b interval. Scalar subdivision reduces the coefficient relaxation while preserving one shared (t,b). Unbounded components and unfinished subdivision branches retain an unresolved status. Appendix F specifies the complete inversion and calibration.

\textbf{Theorem 5. Ancestor-profile validity.} Allocate \(\alpha_A\) to order qualification, \(\alpha_F\) to the regression confidence set, and \(\alpha_u\) and \(\alpha_x\) to the two residual variance comparisons. Whole-domain exclusion reports a correct edge except on an event of probability at most

\[
\alpha_A+\alpha_F+\alpha_u+\alpha_x. \tag{14}
\]

At a true missing row the actual b belongs to \(C_{b}\) on its covering event, and the actual t satisfies both variance-ratio comparisons on their acceptance events. Every corresponding coefficient branch survives outer elimination. Complete exclusion is therefore contained in the union of the four error events. Shared control observations and reuse between stages are handled by this event inclusion. For a predeclared or simultaneously covered query universe Q, the graph-family bound is

\[
\begin{aligned}&\Pr\{\text{at least one false report}\}\\
&\quad\le\alpha_A+\sum_{q\in Q}(\alpha_{F,q}+\alpha_{u,q}+\alpha_{x,q}).\end{aligned}
\]

Here \(\alpha_A\) covers the qualification event. With M Monte Carlo draws, a finite component cutoff requires \(\alpha_r\ge1/(M+1)\). Appendix G gives the calibration-resolution and power conditions.

In the experiment, \(\alpha_A\)=1/100 is divided across three variance-shift ancestry tests, \(\alpha_F\)=1/50, and \(\alpha_u\)=\(\alpha_x\)=1/100. An observed variance change at coordinate i under target j certifies that j is an ancestor of i under (1). Tests for (u,X), (x,V) and (v,Y) supply the required order. The total single-query bound is 1/20.

\textbf{Proposition 4. Uniform profile separation.} On the compact parameter class of Proposition 3,

\[
\inf_t\frac{\max\{|\Delta_u\operatorname{Var}R(t,b_0)|,
|\Delta_x\operatorname{Var}R(t,b_0)|\}}{1+t^2}\ge c_*s
\]

for a uniform \(c_*>0\). The Fieller radius includes covariance error, finite test width and the square root of its quadratic-coefficient enclosure error. With resolvable component levels, \(\eta_F\le c_1s^2\), \(\eta_V\le c_2s\) and sufficient subdivision, the procedure attains the fixed-dimensional \(s^{-2}\) sample order. Appendix G gives the complete contraction and perturbation bounds. Proposition 3 supplies the matching lower order.

\textbf{Algorithm 2. Qualified ancestor-profile certificate.}

\begin{enumerate}
\def\labelenumi{\arabic{enumi}.}
\tightlist
\item
  Qualify the complete parent domain and certify the common order with its declared error allocation.
\item
  Enclose the within-cell centered scatters and compute every component of the regression confidence set \(C_{b}\).
\item
  For each bounded component, intersect the four quadratic outer sublevel sets over the complete real t line.
\item
  Subdivide a retained b interval while the numerical budget permits. Preserve unbounded and unfinished branches.
\item
  Report the edge when every component is excluded, and otherwise return an unresolved query.
\end{enumerate}

\section{Experiments}\label{experiments}

The Gaussian experiment evaluates the ancestor-profile certificate and its confidence representations. The connected population experiment evaluates common-DAG support, target coverage and parent-domain decisions. Both use the final specifications and seed rules fixed before the reported replay. Appendix D evaluates the four measured component estimators on the public acquisitions used in their development. The reproducibility package contains the final experiment specifications, inputs, code and outputs.

\subsection{Gaussian power and confidence-region geometry}\label{gaussian-power-and-confidence-region-geometry}

The Gaussian covariance experiment uses four parameter templates, gaps 0,1/8,1/4,1/2, four sample budgets and 32 independent replicates per template-gap-budget combination. The 2048 conditions comprise 512 null and 1536 positive cases. For template index iota in \{0,1,3,4\} and zero-based coordinates r,s, every baseline is I+H with

\[
H_{rr}=(-1)^{\iota}/128,\qquad
H_{rs}=(-1)^{\iota+r+s}/96\quad(r\ne s).
\]

Every procedure receives the same Gaussian sample scatter. Experiment sources are independent across conditions, and each budget uses a separate calibration bank. Appendix I specifies the design.

Three references use common-DAG rectangles, those rectangles with the same certified order, and one Fieller component box. The rectangle comparisons change the confidence-region construction. The single-box and subdivision procedures share their Fieller set, variance constraints and error budget, isolating numerical relaxation. Different box bounds can be attained at incompatible nuisance values. Subdivision tightens their joint enclosure.

\par

\noindent

\begin{table*}[t]
\begin{minipage}[t]{.49\linewidth}
\textbf{Table 2. Ancestor-profile positive shortcut reports, with 128
independent conditions per entry.}

\begin{center}
\small
\begin{tabular}{@{}p{\dimexpr0.230000\linewidth-2\tabcolsep\relax}p{\dimexpr0.180000\linewidth-2\tabcolsep\relax}p{\dimexpr0.230000\linewidth-2\tabcolsep\relax}p{\dimexpr0.180000\linewidth-2\tabcolsep\relax}p{\dimexpr0.180000\linewidth-2\tabcolsep\relax}@{}}
\toprule
\textbf{Gap s} & \textbf{8192 rows} & \textbf{32768 rows} & \textbf{131072 rows} & \textbf{524288 rows} \\
\midrule
1/8 & 0/128 & 0/128 & 13/128 & 98/128 \\
1/4 & 0/128 & 1/128 & 61/128 & 127/128 \\
1/2 & 0/128 & 2/128 & 118/128 & 128/128 \\
\bottomrule
\end{tabular}
\end{center}
\end{minipage}\hfill
\begin{minipage}[t]{.49\linewidth}
\textbf{Table 3. Joint-reference comparison on the same 2048
conditions.}

\begin{center}
\small
\begin{tabular}{@{}p{\dimexpr0.333333\linewidth-2\tabcolsep\relax}p{\dimexpr0.333333\linewidth-2\tabcolsep\relax}p{\dimexpr0.333333\linewidth-2\tabcolsep\relax}@{}}
\toprule
\textbf{Outcome} & \textbf{Scalar interval procedure} & \textbf{Joint polynomial reference} \\
\midrule
Positive edge reports & 548 & 548 \\
Outer systems retaining coefficients & 1474 & 1474 \\
Unqualified conditions & 26 & 26 \\
Pairwise report disagreements & 0 & 0 \\
\bottomrule
\end{tabular}
\end{center}
\end{minipage}
\end{table*}
\par\medskip

The three references report zero positive edges at every listed setting. Relative to the same-constraint single-box reference, subdivision supplies all 548 positive reports. All 512 null conditions produce zero reports. At the smallest budget the order qualifies in 121/128 null cases and in 125/128,122/128,118/128 positive cases at the three increasing gaps. Every condition at higher budgets qualifies. Across all conditions, 548 reports are certified, 1431 retain a feasible outer profile point, 43 retain an unbounded Fieller component and 26 remain unqualified. The remaining cases show how ancestry qualification and the Fieller-set geometry govern finite-budget power.

Figure S1 gives template-specific power curves and pointwise 95 percent Wilson intervals conditional on each budget's calibration bank. Finite-sample error control averages over calibration and experiment randomness. Detection increases with both sample size and gap, in agreement with the separation margin in Proposition 4.

A paired joint reference uses the same 2048 inputs, outward Fieller components and variance constraints. It decides the rational outer-polynomial system in Appendix J, with a shared nuisance pair. \citep{ref25}

\par

\noindent

\par\medskip

The 1431 midpoint-retained and 43 unbounded-component conditions admit coefficients in this outer system. Its 2022 qualified decisions agree with scalar subdivision. These are paired computational checks on the power study's inputs. Completing a retained coefficient pair to a DAG requires the additional response-column constraints.

\subsection{Connected support, target coverage and parent domains}\label{connected-support-target-coverage-and-parent-domains}

Two downstream families instantiate Theorem 3. One uses successive width-two layers with repeated complete-bipartite connections. The other has a chain plus links three coordinates ahead. Each downstream coefficient is 1/4 and w connects to the first two added nodes with coefficient 1/64. Both row and column absolute sums of D are at most 1/2. The complete design has 16,32,64 nodes, four core templates and four gap states, giving 96 population conditions.

\par

\noindent

\begin{table*}[t]
\textbf{Table 4. Exact connected-family identification and query
results.}

\begin{center}
\small
\begin{tabular}{@{}p{\dimexpr0.080000\linewidth-2\tabcolsep\relax}p{\dimexpr0.250000\linewidth-2\tabcolsep\relax}p{\dimexpr0.190000\linewidth-2\tabcolsep\relax}p{\dimexpr0.250000\linewidth-2\tabcolsep\relax}p{\dimexpr0.230000\linewidth-2\tabcolsep\relax}@{}}
\toprule
\textbf{Nodes} & \textbf{Positive conditions with unique complete support} & \textbf{Null conditions retaining both supports} & \textbf{Missing models after one core target is removed} & \textbf{Correct complete-target parent-domain decisions} \\
\midrule
16 & 24/24 & 8/8 & 96/96 & 96/96 \\
32 & 24/24 & 8/8 & 96/96 & 96/96 \\
64 & 24/24 & 8/8 & 96/96 & 96/96 \\
\bottomrule
\end{tabular}
\end{center}
\end{table*}
\par\medskip

Every target-factor set is recovered exactly. Each positive condition has the two compatible coefficient matrices in Theorem 3 and one complete support. Each zero-gap condition retains the physical chain and its alternative shortcut support. Removing any one core target gives a compatible acyclic missing-shortcut witness. Enlarging the query's remaining parent domain from 2 to 4 to 8 coordinates preserves every expected decision, including the zero coefficients on added downstream parents. The unchanged query decision across these enlarged domains follows the block-triangular response equations in Appendix E.

The connected experiment evaluates population support and target coverage. The sampled experiment evaluates the core query. Theorem 3 links them through a connected construction whose core marginal preserves the sampling law used in Section 6.1.

\subsection{Measured intervention designs}\label{measured-intervention-designs}

Standard light A1 has eight targeted controls among ten known roots and nine untargeted readouts. Proposition 2 characterizes its full-parent interface. A terminal-readout restriction adds physical parent information. Wind and camera require sampling treatment for sequential controls and AR trajectories. The optical catalogue requires a common-regime and pixel-target mapping. These interfaces motivate the application-specific response and conditional-coefficient analyses in Appendix D. That appendix retains their complete fixed-method results and reference definitions. \citep{ref11,ref12}

\section{Discussion}\label{discussion}

In the constructed families, common acyclicity identifies the shortcut, and a response-degree condition identifies complete support while coefficients remain ambiguous. Connected extensions preserve this core information as dimension grows. Finite candidate compatibility turns the identification question into bounded-parent decisions. Certified ancestry then supplies the observed cross-moment needed for scalar Gaussian confidence inversion.

The numerical results separate identifying information from its confidence representation. Removing core targets retains explicit missing-edge DAGs. With the targets present, scalar subdivision recovers the decisions of the joint polynomial confidence system and improves on its single-box relaxation. Proposition 2 characterizes why experiments targeting only roots require additional parent information to certify most direct root-to-readout edges. A measured core application calls for informative internal targets or source-supported parent restrictions together with a matching sampling model.

%% file: appendices.tex
\section{Experimental design}\label{experimental-design}

The study has three parts. The Gaussian power experiment crosses four parameter templates, four gap states, four sample budgets and 32 replicate seeds. The connected population experiment crosses two graph structures, three sizes, four templates and four gap states. The measured study uses the acquisitions in Table D5 with the four fixed estimators in Table D2.

The four core templates use the following parameters. In the power experiment, s ranges over 0,1/8,1/4,1/2. The construction and the coefficient alternatives are those of Section 4.

\begin{center}
\small
\begin{tabular}{@{}p{\dimexpr0.160000\linewidth-2\tabcolsep\relax}p{\dimexpr0.090000\linewidth-2\tabcolsep\relax}p{\dimexpr0.090000\linewidth-2\tabcolsep\relax}p{\dimexpr0.090000\linewidth-2\tabcolsep\relax}p{\dimexpr0.170000\linewidth-2\tabcolsep\relax}p{\dimexpr0.200000\linewidth-2\tabcolsep\relax}p{\dimexpr0.200000\linewidth-2\tabcolsep\relax}@{}}
\toprule
\textbf{Index} & \textbf{a} & \textbf{c} & \textbf{b} & \textbf{k} & \textbf{l} & \textbf{e} \\
\midrule
0 & 1/2 & 3/8 & 1 & 5/8 & 1/4 & -3/4 \\
1 & 5/8 & 3/8 & 5/4 & 3/4 & 1/4 & -1 \\
3 & 1/2 & 1/2 & 1 & 7/8 & 1/4 & -3/2 \\
4 & 5/8 & 1/2 & 5/4 & 3/4 & 1/4 & -3/4 \\
\bottomrule
\end{tabular}
\end{center}

\par

\noindent

\begin{minipage}{\linewidth}
\textbf{Table A1. Statistical settings of the Gaussian confidence-region
comparisons.}

\begin{center}
\small
\begin{tabular}{@{}p{\dimexpr0.333333\linewidth-2\tabcolsep\relax}p{\dimexpr0.333333\linewidth-2\tabcolsep\relax}p{\dimexpr0.333333\linewidth-2\tabcolsep\relax}@{}}
\toprule
\textbf{Procedure} & \textbf{Error allocation} & \textbf{Confidence representation} \\
\midrule
Common-DAG rectangles & \(\alpha_\Sigma=1/20\) & Covariance regions from (10), followed by target-factor and common-DAG
constraints \\
Ordered rectangles & \(\alpha_\Sigma=1/25\), \(\alpha_A=1/100\) & The covariance regions contracted by the same certified order used by
the scalar procedure \\
One Fieller box & \(\alpha_A=1/100\), \(\alpha_F=1/50\), \(\alpha_u=\alpha_x=1/100\) & Every Fieller component is processed once, with the four shared variance
inequalities \\
Scalar interval procedure & Same allocation as one Fieller box & Adaptive scalar subdivision with a 2048-interval budget \\
\bottomrule
\end{tabular}
\end{center}
\end{minipage}
\par\medskip

Each procedure has total error budget \(1/20\) for its declared query. The order-qualified rectangle allocation includes the order event in this total. The one-box and scalar procedures share their calibration banks, cutoffs, Fieller components, variance constraints and unbounded-component rule. Disabling subdivision gives the one-box procedure. Both retain an unresolved query whenever their outer constraints remain feasible.

The power procedure uses M=1023 calibration draws per sample budget and the component allocation in Theorem 5. The three confidence-region references receive the same generated covariance input as the scalar procedure. The joint polynomial reference receives the same Fieller and variance constraints. Appendix I lists all 64 template-gap-budget settings, each with 32 replicates. Table D4 lists every optical acquisition in the final linear specification.

\section{Numerical covariance enclosures}\label{numerical-covariance-enclosures}

For the finite input values, append a column of ones and denote its exact augmented Gram matrix by G. Let Ghat contain the floating dot products and Ahat the floating absolute-product sums. Let \(u_{\rm fp}=2^{-53}\) be the binary64 unit roundoff. The declared IEEE dot-product model uses the explicit sample-count product

\[
\gamma_n=\frac{n\,u_{\rm fp}}{1-n\,u_{\rm fp}},\qquad n\,u_{\rm fp}<\tfrac12.
\]

The product-below-one-half gate ensures \(\gamma_{n}\)\textless1 for the absolute-sum reconstruction. For an outward absolute-underflow allowance \(\tau_{\mathrm{uf}}\), enclose the exact absolute-product sum by

\[
A_{jk}^+=\frac{\widehat A_{jk}+\tau_{\rm uf}}{1-\gamma_n},\qquad
G_{jk}\in\widehat G_{jk}\mathbin{\pm}(\gamma_nA_{jk}^++\tau_{\rm uf}).
\]

The declared allowance is an upper enclosure of \(2n\,2^{-1074}/(1-\gamma_n)\). It covers subnormal operations under the stated dot-product model. These expressions use outward endpoints. Each sample covariance is enclosed by evaluating

\[
\{G_{jk}-G_{0j}G_{0k}/n\}/(n-1)
\]

with the resulting intervals. Finite-input overflow and an unavailable dot-product bound retain an unresolved status. This numerical enclosure concerns the actual observation values supplied to the sampling interface.

The factor boxes contain the full normalized population candidates. Their mandatory support is an inner graph constraint within an outer numerical region. Removing a row candidate requires a whole-interval exclusion. A retained branch supplies an unresolved finite-sample query. An exact population branch that passes the complete decision supplies an explicit compatible model.

\section{Continuous separation margins}\label{continuous-separation-margins}

Fix positive lower and finite upper parameter bounds, a positive lower bound on k-c, an upper bound \(kl\le\tau<1\) and \(0\le s\le s_{\max}\). In the u contrast, the two factors differ by (b-e)ac+s in coordinate w. In the x contrast, they differ by k-c in coordinate v. In the v contrast they differ by l in coordinate x. These three separations and the target anchors have uniform positive lower bounds. The factor map is therefore uniformly Lipschitz near this compact class.

For a missing row in the physical-v and physical-x branch, \(\beta_{v}\)=b and \(\beta_{x}\)=0. Its two u discrepancies are s and (e-b)ac. The second stays uniformly separated from zero. With alternate x and physical v, \(\beta_{v}\)=b and \(\beta_{x}\)=(e-b)k, and the two u discrepancies are s+(b-e)ak and (e-b)a(c-k), both uniformly positive. Alternate v forces the mandatory x-v response cycle. Finitely many strict inequalities therefore supply a common sufficient precision \(\epsilon\le Cs\) for s\textgreater0.

The covariance spectra in (8) stay in {[}1/4,7/4{]}. On the same compact parameter range, the maximum-row-sum normalization is Lipschitz. The covariance difference between the s and zero experiments has norm bounded by a constant times s. Gaussian KL is then bounded by a constant times n \(s^{2}\). Choosing the physical chain at zero and the fixed strong representative at positive s establishes the lower sample order. The sufficient and necessary orders concern fixed-dimensional experiments and fixed nontrivial testing errors.

\section{Measured interfaces, references and complete comparisons}\label{measured-interfaces-references-and-complete-comparisons}

\textbf{Response components and measured results.}

Known acquisition laws also give tests of root-to-readout intervention-response relations. Under a sharp root-absence null, target reassignment leaves the indexed nuisance and response vector invariant. Target-blind residual adjustment can therefore improve a randomization statistic while preserving its null rank distribution. We use within-acquisition permutation ranks for standard light, exact binary-assignment KS tests for wind, and full-trajectory source redraws for the camera. A Gaussian conditional response model gives pixel-group F tests for the optical device. Appendix D specifies the four statistics and their identifying interfaces.

For an inherited graph \(\widehat E_0\) and certified additions \(\widehat C\), the final estimator is \(\widehat E_0\cup\widehat C\). Correct additions increase true positives and decrease false negatives while preserving the inherited false-positive count. The optical application combines MuDo-CoM with nominal conditional-row additions. Appendix H gives the randomization and conditional-regression guarantees.

The four estimators were developed on the public acquisitions evaluated here. The final estimators, coordinates, budgets, seed rules and scoring maps were fixed before the reported replay. The comparisons describe fixed-method performance on these acquisitions, with nominal raw P values from Appendix H. That appendix gives error control for prespecified and null-orbit-invariant rules.

Standard light, wind and camera acquisitions come from Causal Chambers. The optical catalogue has thirteen acquisitions and 13000 rows, including a purely confounded regime. Table D2 specifies their assignment, noise and reference-coordinate interfaces. \citep{ref11,ref12}

The retained external panel comprises PC, GES, GIES, Gaussian HC-BIC, DirectLiNGAM, NOTEARS, DAGMA, GOLEM, MuDo-CoM and the authors' WVM implementation. Common task information and the stated graph adapters are supplied according to each acquisition interface. \citep{ref2,ref4,ref13,ref14,ref15,ref16,ref17,ref18,ref19,ref20}

\par

\noindent

\begin{minipage}{\linewidth}
\textbf{Table D1. Working-reference response-relation scores in four
measured configurations.}

\begin{center}
\small
\begin{tabular}{@{}p{\dimexpr0.360000\linewidth-2\tabcolsep\relax}p{\dimexpr0.110000\linewidth-2\tabcolsep\relax}p{\dimexpr0.220000\linewidth-2\tabcolsep\relax}p{\dimexpr0.130000\linewidth-2\tabcolsep\relax}p{\dimexpr0.180000\linewidth-2\tabcolsep\relax}@{}}
\toprule
\textbf{Configuration and declared interface} & \textbf{Estimator F1} & \textbf{Strongest fitted comparator} & \textbf{Comparator F1} & \textbf{Difference in percentage points} \\
\midrule
Standard light A1, randomized roots and hidden blue & 0.836364 & GIES & 0.769231 & 6.71 \\
Standard wind, source-qualified refinement & 1.000000 & GIES & 0.810811 & 18.92 \\
Camera, numeric readouts and three acquisition conditions & 1.000000 & GIES & 0.984615 & 1.54 \\
Optical device, all thirteen acquisitions, linear response slots & 0.908213 & DirectLiNGAM & 0.873096 & 3.51 \\
\bottomrule
\end{tabular}
\end{center}
\end{minipage}
\par\medskip

\emph{The four rows use application-specific estimators. The first three references contain unknown physical labels. The camera/GIES F1 difference is \(1/65\) on its working reference and ranges over \([-1/65,+1/65]\) across compatible completions.}

Optical micro F1 pools the thirteen acquisition-level pixel-to-sensor comparisons. Its macro F1 is 0.836763, compared with GIES at 0.755090, and 9/13 acquisitions recover their reference slots.

The optical reference contains 108 positive slots among 117. Reporting all slots gives micro F1 0.960000 and macro F1 0.923077, with nine false positives and zero false negatives. The row estimator has precision 0.9495 and recall 0.8704, compared with 0.9231 and 1 for this all-slot reference.

The optical estimator's nominal group-F additions contribute 27 correct reference edges to MuDo-CoM. Its final graph has 94 true positives, five false positives and fourteen false negatives. The five inherited errors occur in the purely confounded acquisition. This separates the gain from row testing from the inherited graph's remaining errors.

\subsection{Acquisition and query interfaces}\label{acquisition-and-query-interfaces}

The acquisition-level source law supplies the root roles. Standard light, wind and camera evaluate root-to-readout intervention-response relations under their documented assignment laws. Rejection concerns the sharp null of no response to the assigned root. The optical application evaluates conditional pixel coefficients with all nine pixel groups as its complete feature-parent domain. Table D2 specifies the four interfaces. \citep{ref11,ref12}

\par

\noindent

\begin{minipage}{\linewidth}
\textbf{Table D2. Source-qualified measured interfaces.}

\begin{center}
\small
\begin{tabular}{@{}p{\dimexpr0.170000\linewidth-2\tabcolsep\relax}p{\dimexpr0.220000\linewidth-2\tabcolsep\relax}p{\dimexpr0.320000\linewidth-2\tabcolsep\relax}p{\dimexpr0.290000\linewidth-2\tabcolsep\relax}@{}}
\toprule
\textbf{Configuration} & \textbf{Acquisitions and measured rows} & \textbf{Row component} & \textbf{Calibration and query allocation} \\
\midrule
Standard light A1 & 11 acquisitions, 20000 rows & Adjusted residual mean/variance evidence and retained-Fisher rank
combination & 32767 synchronized assignment draws, 90 pairs, level 1/1800 per pair \\
Standard wind & 28 acquisitions, 23449 rows & Exact randomization KS with outcome-only pressure centering selection & 308 acquisition/readout tests, level 1/6160 each, union into 231
coordinate slots \\
Camera & Three AR conditions, 4096 rows each & Nuisance-adjusted expanded-feature trajectory randomization & 32767 source draws, 50 pairs, three-condition union, level 1/3000 per
condition/pair \\
Optical device & 13 acquisitions, 1000 rows each & Nominal conditional pixel-group F additions to MuDo-CoM & Nine groups per acquisition, original P threshold 0.05/9, separate
acquisition families \\
\bottomrule
\end{tabular}
\end{center}
\end{minipage}
\par\medskip

For A1, the ten observed controls are red, green, \texttt{pol\_1}, \texttt{pol\_2}, \texttt{\ensuremath{l_{11}}}, \texttt{\ensuremath{l_{12}}}, \texttt{\ensuremath{l_{21}}}, \texttt{\ensuremath{l_{22}}}, \texttt{\ensuremath{l_{31}}} and \texttt{\ensuremath{l_{32}}}. Blue is hidden. The nine readouts are current, \texttt{angle\_1}, \texttt{angle\_2}, \texttt{ir\_1}, \texttt{ir\_2}, \texttt{ir\_3}, \texttt{vis\_1}, \texttt{vis\_2} and \texttt{vis\_3}. The 90 ordered slots are their Cartesian product. Within each acquisition the target-blind nuisance adjustment uses the other controls. Under the sharp root-absence null, the complete indexed nuisance/readout vector is invariant under the permitted reassignment of the target vector.

Scale the other A1 controls to {[}-1,1{]}. The nuisance design has an intercept, each control's first three powers and all pairwise products. Ordinary least squares gives R, centered within acquisition. The target features are its scaled linear coordinate and square, with sine and cosine of twice the angle added for polarizers. Center and unit-normalize every feature using the sorted target multiset. Center and unit-normalize both R and its square. The retained local joint score is the maximum absolute inner product of any target feature with either normalized response feature. Constant columns contribute zero. This defines the mean/variance statistic and keeps its normalization invariant throughout the target permutation orbit.

For observed-plus-reference row b and acquisition g, let \(r_{bg}\) be the inclusive upper rank of that statistic among N=32768 rows. All eleven acquisition ranks are combined by

\[
F_b=\prod_{g=1}^{11}r_{bg},\qquad
L_b=\mathbf1\{11\min_g r_{bg}\le Nq\},\quad q=1/1800.
\]

Order rows first by \(L_{b}\)=1 and then by smaller \(F_{b}\) within each part. Its inclusive rank divided by N is the original combined P. Each acquisition contributes at most \(\lfloor Nq/11\rfloor\) rows to \(L\), so at most \(\lfloor Nq\rfloor\) rows have \(L=1\). This ordering retains that original rejection event and is equivariant in the synchronized reference rows. For a rule satisfying Appendix H's selection condition, source-assignment exchangeability gives its family bound. Acquisition scores can have shared disturbances. These 20000 rows also form the complete external-baseline inputs.

The wind controls are hatch, \texttt{load\_in}, \texttt{load\_out}, \texttt{osr\_1}, \texttt{osr\_2}, \texttt{osr\_ambient}, \texttt{osr\_downwind}, \texttt{osr\_in}, \texttt{osr\_intake}, \texttt{osr\_mic}, \texttt{osr\_out}, \texttt{osr\_upwind}, \texttt{pot\_1}, \texttt{pot\_2}, \texttt{res\_in}, \texttt{res\_out}, \texttt{\ensuremath{v_{1}}}, \texttt{\ensuremath{v_{2}}}, \texttt{\ensuremath{v_{in}}}, \texttt{\ensuremath{v_{mic}}} and \texttt{\ensuremath{v_{out}}}. Its eleven readouts are \texttt{current\_in}, \texttt{current\_out}, \texttt{rpm\_in}, \texttt{rpm\_out}, mic, \texttt{signal\_1}, \texttt{signal\_2}, \texttt{pressure\_upwind}, \texttt{pressure\_intake}, \texttt{pressure\_ambient} and \texttt{pressure\_downwind}. Each acquisition supplies a binary randomized flag and one controlled target. The 308 tests retain the actual observed allocation counts and inclusive ties. For the four pressure readouts and \(n\ge2\), define \(K=\max\{1,\lfloor n/50\rfloor\}\) and partition acquisition positions into K near-equal contiguous blocks. Let P and \(P_{0}\) denote block and global centering, and put \(d=(n-K)/(n-1)\). Use the block-centered outcome when its pooled energy satisfies

\[
\|PY\|^2<d^2\|P_0Y\|^2.
\]

When \(K=1\), \(P=P_0\) and \(d=1\). The strict selector inequality is an equality, so this case uses the raw outcome. The other outcome branch, including ties, uses the raw values. This selector is fixed throughout the null assignment orbit because it uses outcomes and positions. Exact two-sample KS upper-tail probabilities are compared with 1/6160. A root/readout slot is reported when one of its acquisition tests rejects. The remaining seven readouts use the raw branch.

The camera roots are red, green, blue, \texttt{pol\_1} and \texttt{pol\_2}. The ten readouts are current, \texttt{angle\_1}, \texttt{angle\_2}, \texttt{ir\_1}, \texttt{ir\_2}, \texttt{ir\_3}, \texttt{vis\_1}, \texttt{vis\_2}, \texttt{vis\_3} and im. The im coordinate is the mean of every supplied 64-by-64 RGB frame with integer channel weights 299,587,114, divided by 1000 times its pixel count. Each analyzed 4096-position prefix comes from the length-30000 archived AR source map, including its complete-trajectory min/max normalization and command quantization. A reference redraw generates that complete target trajectory. The source-law qualification fixes its recursion and hardware command map before response fitting.

For each camera condition, the nuisance design contains an intercept, degree-one and degree-two monomials of the other four roots, and 16 fixed position blocks. The target features contain its scaled linear coordinate and square. Polarizers also have sine and cosine of twice their angle. With nuisance residual R and centered target features \(f_{k}\), the statistic is

\[
T=\max_k\frac{|\langle f_k-\bar f_k,R\rangle|}{\|f_k-\bar f_k\|_2\|R\|_2}.
\]

A zero denominator contributes zero. Inclusive source-reference ranks give condition-specific original P values. The pair value is \(\min\{1,3\min_c p_c\}\) and the pair threshold is 0.001, yielding the stated 150-test allocation. Every external method receives the same 12288 original rows, fifteen coordinates, three condition labels and root/readout domain.

Each optical acquisition provides nine pixels, a sensor and, in eleven-column files, a recorded LED command. The ten-column image-section acquisition uses an intercept nuisance. In the primary working model,

\[
Y=\beta^\top\phi(X)+h(Z)+\varepsilon,\qquad
\varepsilon\mid X,Z\sim N(0,\sigma^2I),\quad\phi_i(X)=X_i/255.
\]

Let C contain the recorded LED-level indicators or the intercept, and choose orthonormal contrasts H with \(HC=0\) and \(HH^\top=I\). Transform both pixels and sensor by H. The conditional model then has Gaussian spherical noise and retains \(\beta\). The full transformed design includes its fitted intercept. For pixel group G of \(q_{G}\) coordinates, compare the constrained and full residual sums of squares by

\[
F_G=\frac{(RSS_R-RSS_F)/q_G}{RSS_F/df}\sim F_{q_G,df},
\qquad df=n-\operatorname{rank}(C)-\operatorname{rank}(D_{\rm full}).
\]

Here \(n-\operatorname{rank}(C)\) is the number of orthonormal transformed rows, and \(D_{\mathrm{full}}\) is the complete fitted transformed design. Singular groups are retained unresolved. The primary linear arm has \(q_G=1\). All external methods receive the same transformed rows and feature-parent domain. A pixel is reported when any of its feature edges is present. The primary optical estimator combines MuDo-CoM with the nominal linear group-F additions.

\subsection{Working-reference scores and unknown labels}\label{working-reference-scores-and-unknown-labels}

Let D be the declared directed coordinate domain, P its listed-positive reference subset, and E the reported point graph after the common domain adapter. The working-reference scores in Table D1 are

\[
F_{1,\mathrm{work}}=\frac{2|E\cap P|}{|E|+|P|},\qquad
\mathrm{SHD}_{\mathrm{work}}=|E\setminus P|+|P\setminus E|.
\]

The all-slot reference sets \(E=D\), giving \(TP=|P|\), \(FP=|D|-|P|\), \(FN=0\) and \(F_{1,\mathrm{all}}=2|P|/(|D|+|P|)\). Table D3 applies this rule to each declared domain.

Outputs in D outside P contribute reference disagreements in this point score. Their physical labels retain an unknown status in the first three configurations. The score's reference object is P. For a physical completion Q=P union S, with S contained in the unknown subset, the corresponding score instead is

\[
F_1(Q)=\frac{2(|E\cap P|+|E\cap S|)}{|E|+|P|+|S|}.
\]

The completed analyses retain comparison ranges over these source-compatible completions. For example, the camera candidate/GIES difference ranges from -1/65 to 1/65, while Table D1 reports its working-reference value 1/65. The domain sizes are 90,231,50 and nine per optical acquisition. Their listed-positive counts are 32,42,32 and nine in each of twelve optical acquisitions. Optical experiment 82 has an empty response reference. F1 is one when both the reference and reported graph are empty, and zero when only the reference is empty. Optical micro F1 pools TP,FP,FN across all thirteen acquisitions. Macro F1 averages their thirteen F1 values, including the empty-reference case.

\subsection{Complete native-method comparison and adapters}\label{complete-native-method-comparison-and-adapters}

All ten external methods retain their native fitted outputs and the common declared domain conversion. For root/readout tasks, retain a reported skeleton relation when its endpoints match an allowed root/readout slot, and orient it from the hardware root to the readout. Native directed and undirected endpoints use that same conversion. The measured tasks supply empty mandatory-edge lists. Common scaling uses the complete declared acquisition rows. Optical feature-level edges use the same pixel-group union rule for every method.

PC uses its Gaussian conditional-independence interface with \(\alpha\)=0.05, maximum conditioning size two and qualified background information. GES uses its retained Gaussian score. GIES uses the conditional Gaussian score given the hardware-controlled roots, with the roots supplied as intervention targets in each group. Gaussian HC-BIC restricts response parents to the recorded roots. DirectLiNGAM receives the qualified ancestry map and its child-by-parent output is transposed to the graph convention. NOTEARS uses lambda1=0.1 and weight threshold 0.3. DAGMA uses lambda1=0.03 and threshold 0.3. GOLEM's reported NV fit inherits its EV warm start. MuDo-CoM uses the fitted mixing matrix, row assignment minimizing the reciprocal absolute diagonal of its inverse, and magnitude-ordered acyclic pruning at threshold 0.3. Author WVM retains its original implementation, supplied acquisition groups and recorded point decisions. WVM uses its authors' bootstrap moment estimator with B=200 and a fitted Gamma law for thresholds and original P values. Its primary levels are 1/1800 for A1, \(1/720\) for wind, 0.001 for camera and 0.05 for each optical response. Wind uses the within-acquisition orthogonal contrasts, and the optical interface uses two fixed contiguous blocks of its transformed rows. The reproduction specification gives the native calibration calls and secondary levels.

\par

\noindent

\begin{minipage}{\linewidth}
\textbf{Table D3. Working-reference F1 for fitted methods and the
all-slot reference. Optical columns use the linear arm and all thirteen
acquisitions.}

\begin{center}
\small
\begin{tabular}{@{}p{\dimexpr0.166667\linewidth-2\tabcolsep\relax}p{\dimexpr0.166667\linewidth-2\tabcolsep\relax}p{\dimexpr0.166667\linewidth-2\tabcolsep\relax}p{\dimexpr0.166667\linewidth-2\tabcolsep\relax}p{\dimexpr0.166667\linewidth-2\tabcolsep\relax}p{\dimexpr0.166667\linewidth-2\tabcolsep\relax}@{}}
\toprule
\textbf{Method} & \textbf{Standard light} & \textbf{Wind} & \textbf{Camera} & \textbf{Optical micro} & \textbf{Optical macro} \\
\midrule
PC & 0.612245 & 0.509804 & 0.703704 & 0.625000 & 0.536108 \\
GES & 0.603175 & 0.436364 & 0.875000 & 0.829787 & 0.720782 \\
GIES & 0.769231 & 0.810811 & 0.984615 & 0.867347 & 0.755090 \\
Gaussian HC-BIC & 0.686567 & 0.385027 & 0.955224 & 0.861538 & 0.748497 \\
DirectLiNGAM & 0.666667 & 0.462500 & 0.782609 & 0.873096 & 0.749321 \\
NOTEARS & 0.270270 & 0.285714 & 0.476190 & 0.457143 & 0.467855 \\
DAGMA & 0.358974 & 0.392857 & 0.533333 & 0.641509 & 0.608326 \\
GOLEM & 0.439024 & 0.333333 & 0.545455 & 0.729412 & 0.684357 \\
MuDo-CoM & 0.491525 & 0.351515 & 0.666667 & 0.744444 & 0.675241 \\
Author WVM & 0.607143 & 0.294118 & 0.533333 & 0.054054 & 0.123077 \\
All-slot reference & 0.524590 & 0.307692 & 0.780488 & 0.960000 & 0.923077 \\
Application-specific row estimator & 0.836364 & 1.000000 & 1.000000 & 0.908213 & 0.836763 \\
\bottomrule
\end{tabular}
\end{center}
\end{minipage}
\par\medskip

The last row uses the four distinct interfaces in Table D2. The optical linear inherited MuDo-CoM graph has TP=67, FP=5 and FN=41. The row additions contribute 27 correct reference edges and zero new false edges, yielding TP=94, FP=5 and FN=14. The five inherited false edges remain in experiment 82.

\par

\noindent

\begin{minipage}{\linewidth}
\textbf{Table D4. Every optical acquisition in the primary inherited
linear result.}

\begin{center}
\small
\begin{tabular}{@{}p{\dimexpr0.160000\linewidth-2\tabcolsep\relax}p{\dimexpr0.090000\linewidth-2\tabcolsep\relax}p{\dimexpr0.090000\linewidth-2\tabcolsep\relax}p{\dimexpr0.090000\linewidth-2\tabcolsep\relax}p{\dimexpr0.170000\linewidth-2\tabcolsep\relax}p{\dimexpr0.200000\linewidth-2\tabcolsep\relax}p{\dimexpr0.200000\linewidth-2\tabcolsep\relax}@{}}
\toprule
\textbf{Experiment} & \textbf{TP} & \textbf{FP} & \textbf{FN} & \textbf{F1} & \textbf{Added correct} & \textbf{Added false} \\
\midrule
69 & 9 & 0 & 0 & 1.000000 & 3 & 0 \\
70 & 6 & 0 & 3 & 0.800000 & 0 & 0 \\
71 & 9 & 0 & 0 & 1.000000 & 1 & 0 \\
72 & 9 & 0 & 0 & 1.000000 & 4 & 0 \\
73 & 9 & 0 & 0 & 1.000000 & 2 & 0 \\
74 & 2 & 0 & 7 & 0.363636 & 0 & 0 \\
75 & 9 & 0 & 0 & 1.000000 & 1 & 0 \\
77 & 9 & 0 & 0 & 1.000000 & 4 & 0 \\
78 & 9 & 0 & 0 & 1.000000 & 4 & 0 \\
79 & 5 & 0 & 4 & 0.714286 & 4 & 0 \\
80 & 9 & 0 & 0 & 1.000000 & 4 & 0 \\
81 & 9 & 0 & 0 & 1.000000 & 0 & 0 \\
82 & 0 & 5 & 0 & 0.000000 & 0 & 0 \\
\bottomrule
\end{tabular}
\end{center}
\end{minipage}
\par\medskip

\clearpage
\subsection{Complete acquisition inventory}\label{complete-acquisition-inventory}

\par

\noindent

\textbf{Table D5. Retained public acquisition identifiers and sample
counts.}

\begingroup
\small
\begin{longtable}{@{}p{\dimexpr0.180000\linewidth-2\tabcolsep\relax}p{\dimexpr0.650000\linewidth-2\tabcolsep\relax}p{\dimexpr0.170000\linewidth-2\tabcolsep\relax}@{}}
\toprule
\textbf{Configuration} & \textbf{Acquisition} & \textbf{Rows} \\
\midrule
\endfirsthead
\toprule
\textbf{Configuration} & \textbf{Acquisition} & \textbf{Rows} \\
\midrule
\endhead
\midrule
\multicolumn{3}{r}{Continued on next page}\\
\endfoot
\bottomrule
\endlastfoot

Standard light & \texttt{uniform\_reference} & 10000 \\
Standard light & \texttt{uniform\_red\_mid} & 1000 \\
Standard light & \texttt{uniform\_red\_strong} & 1000 \\
Standard light & \texttt{uniform\_green\_mid} & 1000 \\
Standard light & \texttt{uniform\_green\_strong} & 1000 \\
Standard light & \texttt{uniform\_l\_11\_mid} & 1000 \\
Standard light & \texttt{uniform\_l\_12\_mid} & 1000 \\
Standard light & \texttt{uniform\_l\_21\_mid} & 1000 \\
Standard light & \texttt{uniform\_l\_22\_mid} & 1000 \\
Standard light & \texttt{uniform\_l\_31\_mid} & 1000 \\
Standard light & \texttt{uniform\_l\_32\_mid} & 1000 \\
Wind & \texttt{validate\_hatch\_mic.csv} & 4000 \\
Wind & \texttt{validate\_hatch\_pressures.csv} & 50 \\
Wind & \texttt{validate\_hatch\_rpms.csv} & 200 \\
Wind & \texttt{validate\_load\_in.csv} & 50 \\
Wind & \texttt{validate\_load\_in\_current\_out.csv} & 100 \\
Wind & \texttt{validate\_load\_in\_mic.csv} & 200 \\
Wind & \texttt{validate\_load\_out.csv} & 49 \\
Wind & \texttt{validate\_load\_out\_current\_in.csv} & 100 \\
Wind & \texttt{validate\_load\_out\_mic.csv} & 200 \\
Wind & \texttt{validate\_load\_out\_pressure\_intake.csv} & 3000 \\
Wind & \texttt{validate\_osr\_1.csv} & 500 \\
Wind & \texttt{validate\_osr\_2.csv} & 500 \\
Wind & \texttt{validate\_osr\_ambient.csv} & 8000 \\
Wind & \texttt{validate\_osr\_downwind.csv} & 200 \\
Wind & \texttt{validate\_osr\_in.csv} & 1000 \\
Wind & \texttt{validate\_osr\_intake.csv} & 200 \\
Wind & \texttt{validate\_osr\_mic.csv} & 500 \\
Wind & \texttt{validate\_osr\_out.csv} & 1000 \\
Wind & \texttt{validate\_osr\_upwind.csv} & 200 \\
Wind & \texttt{validate\_pot\_1.csv} & 200 \\
Wind & \texttt{validate\_pot\_2.csv} & 200 \\
Wind & \texttt{validate\_res\_in.csv} & 1000 \\
Wind & \texttt{validate\_res\_out.csv} & 1000 \\
Wind & \texttt{validate\_v\_1.csv} & 200 \\
Wind & \texttt{validate\_v\_2.csv} & 200 \\
Wind & \texttt{validate\_v\_in.csv} & 200 \\
Wind & \texttt{validate\_v\_mic.csv} & 200 \\
Wind & \texttt{validate\_v\_out.csv} & 200 \\
Camera & \texttt{ar\_1\_uniform\_ref} & 4096 \\
Camera & \texttt{ar\_1\_uniform\_pol\_1\_90} & 4096 \\
Camera & \texttt{ar\_1\_uniform\_pol\_12\_30} & 4096 \\
Optical & Experiment 69 & 1000 \\
Optical & Experiment 70 & 1000 \\
Optical & Experiment 71 & 1000 \\
Optical & Experiment 72 & 1000 \\
Optical & Experiment 73 & 1000 \\
Optical & Experiment 74 & 1000 \\
Optical & Experiment 75 & 1000 \\
Optical & Experiment 77 & 1000 \\
Optical & Experiment 78 & 1000 \\
Optical & Experiment 79 & 1000 \\
Optical & Experiment 80 & 1000 \\
Optical & Experiment 81 & 1000 \\
Optical & Experiment 82 & 1000 \\

\end{longtable}
\endgroup

\par\medskip

\section{Connected construction and target-mask witnesses}\label{connected-construction-and-target-mask-witnesses}

\subsection{Model and construction}\label{model-and-construction}

Let the four-node core and its three informative contrasts be those of Theorem 2 in Section 4, with coordinates (u,x,v,w). Write its physical response as \(A_{0}\) and its paired target factors as (\(a_{j}\),\(v_{j}\)). Let D be a strictly lower triangular q-by-q downstream coefficient matrix. Let c be a q-vector with nonzero entries only at declared downstream entry nodes. Put

\[
A_D=(I-D)^{-1},\qquad h=A_Dc,\qquad
L=\begin{bmatrix}I_4\\h e_w^\top\end{bmatrix}.
\]

The connected physical graph has coefficient matrix

\[
B_* = \begin{bmatrix}B_0&0\\c e_w^\top&D\end{bmatrix}
\]

and response

\[
A_* = \begin{bmatrix}A_0&0\\h e_w^\top A_0&A_D\end{bmatrix}.
\]

Lift each original target contrast to \(q_j L S_j L^\top\), retaining its original scale \(q_{j}\). Target w and every downstream coordinate with an informative positive rank-one contrast along its \(A_*\) response column. Such a contrast has one normalized response candidate. Use a baseline diag(\(\Sigma_{\rm core}\),\(I_{q}\)), where \(\Sigma_{\rm core}\) is the non-unit baseline of the Gaussian power experiment. The core marginal of each original target experiment is unchanged.

\subsection{Uniform covariance bounds}\label{uniform-covariance-bounds}

Assume \(\|D\|_2\le\rho<1\) and \(\|c\|_2\le(1-\rho)/16\). Then \(\|h\|_2\le1/16\) and \(\|L\|_2^2=1+\|h\|_2^2\le257/256\). Each original scaled contrast has operator norm at most 3/4. The lifted original cells therefore have eigenvalues in {[}125/1024,1923/1024{]}, uniformly in q. The baseline eigenvalues lie in {[}7/8,9/8{]}. Scale each additional positive rank-one contrast to have maximum absolute row sum 1/4. Its operator norm is at most 1/4, preserving positive definiteness and a dimension-independent upper bound.

\subsection{Complete-support theorem}\label{complete-support-theorem}

Suppose all p=4+q target contrasts are available and w has at most two structural parents. For s\textgreater0 every compatible DAG has the same complete support as \(B_*\). There are exactly two coefficient matrices. They have the same fixed c and D, the same chain coefficients, and shortcut coefficients s and (e-b)ac.

To prove this, each downstream column and the w column is fixed by its rank-one contrast. The three original columns have exactly their lifted factor alternatives. The response matrix of every selection has block form

\[
A=\begin{bmatrix}A_C&0\\h e_w^\top A_C&A_D\end{bmatrix}.
\]

The local x-v cycle excludes alternate v. A compatible acyclic core has \(u<x<v<w\). Block inversion gives

\[
I-A^{-1}=\begin{bmatrix}I-A_C^{-1}&0\\c e_w^\top&D\end{bmatrix}.
\]

Thus the core degree argument excludes alternate x. Both remaining u choices are compatible. Their shortcut coefficients are nonzero for positive s and their downstream structural matrices coincide. This proves the full enumeration. At s=0, the compatible physical chain and the alternative shortcut support remain distinct, with the chain being the unique minimum-cardinality support.

The stable feedback countermodel extends by replacing \(B_{0}\) with \(B_{c}\) in the same block matrix. Its eigenvalues are the core feedback eigenvalues together with zeros from D. The selected lifted alternate factors are mapped to scalar target vectors. The added rank-one targets use the same w and downstream responses in both models. Hence the same complete covariance experiment also has a stable zero-diagonal missing-shortcut representation. The legal residual marginal and joint restrictions remain unchanged for additional downstream targets. Target w is excluded from the invariance requirement for the w structural error.

\subsection{Necessity of the three core targets}\label{necessity-of-the-three-core-targets}

Retain w and every downstream target, and remove one core target. There is a compatible acyclic missing-shortcut model, even under the same response-degree restriction.

\begin{itemize}
\tightlist
\item
  Without target u, keep physical x and v and use the zero-shortcut chain. Its unknown u response may change.
\item
  Without target x, choose physical u and v. Let the unknown x response be \(e_x+(s/a)e_w\). The core structural coefficients are \(B_{xu}=a\), \(B_{vu}=ac\), \(B_{wv}=b\) and \(B_{wx}=s/a\). These coefficients satisfy the known u and v columns with \(B_{wu}=0\).
\item
  Without target v, choose alternate u and x and set the unknown v response to \(e_v+ee_w\). The core coefficients are \(B_{xu}=a\), \(B_{vx}=k\), \(B_{vu}=a(c-k)\) and \(B_{wv}=e\). This model has no shortcut.
\end{itemize}

Lift each model with the same c and D. Block inversion preserves the downstream structural matrix and every retained singleton response column. These are explicit partial-target countermodels. The target mask is therefore an identifying condition that can be evaluated directly.

\subsection{Parent-domain and finite-sample consequences}\label{parent-domain-and-finite-sample-consequences}

With all target columns observed, enlarging the missing-query parent domain from \{x,v\} to include downstream nodes retains the null class correctly. Every downstream column has zero entries in the core rows. The exact parent equations force each added downstream coefficient to zero, after triangular elimination. The query result is unchanged, while its calculation can depend on the declared parent-domain size.

The ancestor-Fieller-quadratic procedure continues to use the core marginal with complete parent domain \{u,x,v\}. Its source law and single-query power are the core law of the independent experiment, because the lifted contrast and block baseline have the same four-dimensional principal submatrices. This inheritance is a consequence of the supplied parent-domain interface. A sampled full-graph recovery claim additionally needs finite-sample coverage of the downstream columns and all corresponding support decisions.

\section{Ancestor covariance, Gaussian inversion and whole-domain elimination}\label{ancestor-covariance-gaussian-inversion-and-whole-domain-elimination}

\subsection{Qualified four-node class}\label{qualified-four-node-class}

Use the centered linear Gaussian SEM with arbitrary correlated errors and the target-row-and-column covariance model. Algorithm 2 uses four independently sampled cells 0,u,x,v, with independent rows within each and a shared control sample across comparisons. The observed coordinates are (U,X,V,Y), with known intervention targets u,x,v and response w. A correct ancestor order \(u<x<v<w\) implies pa(x) is contained in \{u\} and, under a missing u-to-w edge, pa(w) is contained in \{x,v\}. Write

\[
X=aU+\epsilon_x,\qquad
R=Y-tX-bV=\epsilon_w.
\]

The order may be externally qualified or certified with a separate error allocation. For the identifying family, three nonzero variance changes give such a certificate. If the variance of coordinate i changes under a target j, then j is an ancestor of i, because its covariance contrast has diagonal entry 2 \(A_{ij}\) \(h_{ij}\). A missing ancestor forces that diagonal entry to zero.

Tests of variance equality for (u,X), (x,V) and (v,Y) therefore give \(u<x<v<w\) whenever all three reject. Their family error is at most the sum of their declared levels. Gaussian variance-ratio pivots or their independently calibrated log ratios give finite-sample tests. Shared control observations are permitted by the error union.

\subsection{An observed ancestor gives a stronger legal joint restriction}\label{an-observed-ancestor-gives-a-stronger-legal-joint-restriction}

Across control and target v, the structural-error covariance block on \{u,x,w\} is unchanged. Hence the joint covariance of (X,R) is unchanged. The variance of X is also unchanged. Adding the same tX to R shows that

\[
(X,Z(b)),\qquad Z(b)=Y-bV,
\]

has a common bivariate Gaussian covariance across these two cells. This statement removes t from the joint restriction. It uses the certified common order to transfer an error-block restriction to an observed nonroot ancestor.

In the identifying family, the target-v covariance difference satisfies

\[
\Delta_v\operatorname{Cov}(X,Z(b))
=q_v l(b_0-b).
\]

Here \(b_{0}\) denotes the physical chain coefficient. Thus this observed-ancestor joint restriction fixes b=\(b_{0}\) at population. The root-only joint restriction retains b=e. At b=\(b_{0}\) the residual variance equations in the u and x cells have no common t when s\textgreater0, as follows from the four-node enumeration.

\subsection{Exact scalar Fieller confidence set for b}\label{exact-scalar-fieller-confidence-set-for-b}

Use independently sampled Gaussian cells 0 and v. Estimate each mean separately. Let \(H_g\) be the centered X scatter. Let \(d_{Y}\) and \(d_{V}\) be the differences between the two separate within-cell regression slopes of Y on X and V on X. The slope difference for Z(b) is \(d_{Y}\)-b \(d_{V}\).

For each cell let \(Q_{g}\)(b) be the residual sum of squares after regressing Z(b) on X with its own intercept and slope. This is a quadratic polynomial in b. Put \(Q(b)=Q_0(b)+Q_v(b)\) and \(d_f=n_0+n_v-4\). At the true missing-edge coefficient, the shared bivariate covariance gives a common conditional variance and common slope of Z on X. Conditional on the observed X designs,

\[
F(b)=\frac{(d_Y-bd_V)^2}
{(H_0^{-1}+H_v^{-1})Q(b)/d_f}
\sim F_{1,d_f}.
\]

The group-specific fitted means and slopes account for the displayed degrees of freedom. Numerator and pooled residual sum of squares are conditionally independent under the null.

A cutoff \(c_{F}\) with acceptance probability at least 1-\(\alpha_F\) gives

\[
\mathcal C_b=\{b:(d_Y-bd_V)^2
-c_F(H_0^{-1}+H_v^{-1})Q(b)/d_f\le0\}.
\]

All calibration banks in Section 6.1 contain M=1023 independent reference draws. This is a Fieller-type confidence set from scalar quadratic test inversion. \citep{ref24} It can be an interval, two rays, the line or an empty set. Every component is retained. It is a confidence set for the nuisance coefficient at a true missing edge under the qualified order.

Independent Monte Carlo cutoffs can use \(k=\lceil(M+1)(1-\alpha_F)\rceil\) and the kth ordered reference statistic, with an infinite cutoff if k=M+1. Acceptance at the upper numerical enclosure is conservative. The guarantee includes the cutoff randomness. This inherits the classical rank principle.

\subsection{Independent-seed sampling through Gaussian sufficient statistics}\label{independent-seed-sampling-through-gaussian-sufficient-statistics}

The Gaussian covariance comparison uses centered sample scatters. For independent Gaussian observations, \(\nu\) S has the Wishart law with \(\nu=n-1\) and population covariance Sigma. Bartlett construction supplies exactly this ideal sampling distribution from a small lower-triangular matrix. It supplies the joint Gaussian scatter law received by every comparison procedure.

The covariance comparison uses independent scatters for every template-gap-budget-replicate condition. Outward Cholesky and Bartlett products enclose each actual generated scatter. The covariance interface preserves the joint scatter law across entries and the control reuse within each condition.

\subsection{Variance-ratio confidence constraints}\label{variance-ratio-confidence-constraints}

Keep the same Gaussian model, certified order and Fieller confidence set \(C_{b}\). For cells g=u,x, compare the residual variance of R(t,b)=Y-tX-bV with control. Use the sample covariance with \(\nu=n-1\) degrees in each cell. At a true missing coefficient pair, its two sample residual-variance ratios have \(F_{\nu,\nu}\) marginals. Their dependence through control is allowed by the error union.

An independently calibrated cutoff \(c_g\ge1\) for max(F,1/F) gives acceptance probability at least \(1-\alpha_g\). The constraints are

\[
h_g(t,b)-c_gh_0(t,b)\le0,\qquad
h_0(t,b)-c_gh_g(t,b)\le0,
\]

where

\[
h_g(t,b)=S_{XX,g}t^2
-2(S_{XY,g}-bS_{XV,g})t
+S_{YY,g}-2bS_{VY,g}+b^2S_{VV,g}.
\]

Four quadratic inequalities, with one shared t and b, describe the variance confidence profile. Calibration uses the kth upper reference enclosure for \(k=\lceil(M+1)(1-\alpha_g)\rceil\). All comparisons use actual sample-covariance enclosures.

\subsection{Full-line elimination of t}\label{full-line-elimination-of-t}

For a bounded interval J containing b, enclose each quadratic's coefficients A,B,C by outward intervals. For \(t\ge0\), the lower polynomial \(A_L t^2+B_L t+C_L\) is below every allowed quadratic. For \(t\le0\), use \(A_L t^2+B_U t+C_L\). Their sublevel sets, intersected with the corresponding half-lines, form an outer set of every feasible t.

Exact rational discriminants and outward square-root bounds compute each quadratic sublevel set. Linear, constant, repeated-root, disconnected-ray and unbounded cases are retained. Intersect the four outer sublevel sets. An empty intersection excludes every coefficient pair in J times the real line. Otherwise subdivide J or retain it unresolved.

The algorithm covers every Fieller component. Empty \(C_{b}\) excludes the missing model through the Fieller test. An unbounded \(C_{b}\) is retained unresolved in the finite-budget procedure. A subdivision cap retains the unprocessed branches. This yields a finite conservative procedure without a numerical coefficient truncation.

\subsection{Validity and power connection}\label{validity-and-power-connection}

At a true missing edge on the correctly qualified order, the true b is in \(C_{b}\) except on its \(\alpha_F\) event. Its true t satisfies each of the two variance-ratio confidence comparisons except on events of probability at most \(\alpha_u\) and \(\alpha_x\). It therefore survives the appropriate interval branch and every outer quadratic intersection. Complete exclusion can only occur on the union of these events or the ancestor-qualification failure event.

Thus false-report probability is at most \(\alpha_A\)+\(\alpha_F\)+\(\alpha_u\)+\(\alpha_x\). The experimental single-query allocation is 1/100+1/50+1/100+1/100=1/20. Each of the three variance-shift ancestry tests uses \(\alpha_A\)/3. A graph family keeps its shared ancestry allocation and sums its declared row-test allocations.

In the complete-support family, the Fieller set contracts to \(b_{0}\) because the slope contrast in V on X equals a nonzero \(q_{v}\) l/Var(X). At b=\(b_{0}\), the two variance constraints have no common t for s\textgreater0. Their common leading t direction is also excluded for large samples because X has a target-specific variance shift. With a bounded Fieller set, the compact coefficient-direction argument and strict separation give eventual exclusion as samples grow and arithmetic precision is controlled. This is a consistency argument for the declared class. Section 6.1 measures finite-budget power and the retained scalar subdivisions.

\section{Uniform margin, finite test width and calibration resolution}\label{uniform-margin-finite-test-width-and-calibration-resolution}

\subsection{Uniform population margin}\label{uniform-population-margin}

Use a compact parameter class with positive lower bounds on a,c,b,k,l,\textbar e\textbar,k-c and 1-kl. Let \(0<s\le s_{\max}\). The intervention scales \(q_{u}\),\(q_{x}\),\(q_{v}\) have positive lower bounds and finite upper bounds. The baseline covariance has eigenvalues in a fixed positive interval.

At the physical b=\(b_{0}\), the residual R(t,b)=Y-tX-bV has variance contrasts

\[
\Delta_u(t,b_0)=2q_u(s-at)\{(e-b_0)ac-at\},
\]

\[
\Delta_x(t,b_0)=2q_x(-t)\{(e-b_0)k-t\}.
\]

There is a constant \(c_*>0\) uniform on this class such that

\[
\inf_{t\in\mathbb R}
\frac{\max(|\Delta_u(t,b_0)|,|\Delta_x(t,b_0)|)}{1+t^2}
\ge c_*s.
\]

Choose a fixed neighborhood \(|t|\le\delta\) of zero small enough that the second factors in both contrasts stay uniformly away from zero. The two absolute contrasts then dominate constants times \textbar s-at\textbar{} and \textbar t\textbar. The inequality \(s\le|s-at|+a|t|\) gives the desired linear lower bound in that neighborhood. Outside this neighborhood, at s=0 the two zero sets have no common point because k\textgreater c and e-\(b_{0}\)\textless0. At positive s the other root s/a is positive, while the nonzero x root is negative and differs from (e-\(b_{0}\))c.~Compactness of the parameters, s in {[}0,\(s_{\mathrm{max}}\){]} and the real line compactified by its two limits gives a positive bound away from zero. The limiting normalized x contrast is 2\(q_{x}\). Rescaling this fixed bound by \(s_{\mathrm{max}}\) completes the uniform claim.

For b in a bounded neighborhood of \(b_{0}\), both normalized variance contrasts are uniformly Lipschitz in b. Thus a sufficiently narrow interval \(|b-b_0|\le c_1s\) retains a positive normalized separation.

\subsection{Deterministic Fieller contraction}\label{deterministic-fieller-contraction}

Let \(\nu_g\)=\(n_{g}\)-1 and \(\nu_{\rm eff}=(\nu_0^{-1}+\nu_v^{-1})^{-1}\). The population slope contrasts obey \(d_{Y,0}=b_0d_{V,0}\), with \(|d_{V,0}|\ge d_*>0\) uniformly on the compact family. Suppose every sample covariance entry is within \(\epsilon_{\rm cov}\) of its population value. Uniform positive definiteness and a bounded \(b_{0}\) give \(|d_Y-b_0d_V|\le C\epsilon_{\rm cov}\) and \(|d_V|\ge d_*/2\) for sufficiently small \(\epsilon_{\rm cov}\).

At an accepted b, the actual Fieller polynomial gives

\[
|d_Y-bd_V|^2\le C(c_F/\nu_{\rm eff})(1+b^2).
\]

For the implemented outer polynomial, let \(\eta_F\) bound each coefficient enclosure width. Its lower polynomial differs from the actual polynomial by at most C \(\eta_F\)(1+\(b^{2}\)). Thus the outer accepted set obeys

\[
|d_Y-bd_V|\le C\{\sqrt{c_F/\nu_{\rm eff}}+\sqrt{\eta_F}\}(1+|b|).
\]

Write \(|b|\le|b_0|+|b-b_0|\) and absorb the last \(|b-b_0|\) term when the displayed test-and-numerical width is small enough. This proves boundedness of every outer component and

\[
\sup_{b\in\mathcal C_b}|b-b_0|
\le C\{\epsilon_{\rm cov}+\sqrt{c_F/\nu_{\rm eff}}+\sqrt{\eta_F}\}.
\]

The finite F-test width persists when the sample covariance equals the population covariance. The squared numerical term arises from boxing the coefficients of a squared affine contrast. The sufficient precision condition for the coefficient representation is \(\eta_F\le cs^2\). The bound follows the implemented quadratic coefficient enclosure.

\subsection{Variance profile and sample order}\label{variance-profile-and-sample-order}

Let \(\eta_V\) bound numerical coefficient error in the four variance inequalities and let \(\delta_b\) be the retained scalar subdivision width. For bounded b, covariance deviations, ratio-cutoff widths and coefficient errors perturb each normalized residual inequality by at most

\[
C\{\epsilon_{\rm cov}+\max_g(c_g-1)+\eta_V+\delta_b\}.
\]

The uniform population margin \(c_*s\) therefore yields complete exclusion whenever these terms and the Fieller contraction radius are below sufficiently small constants times s. One sufficient precision choice is \(\eta_F\le c_1s^2\) and \(\eta_V\le c_2s\), with scalar intervals narrow enough to preserve the margin. These conditions separate arithmetic accuracy from the sampling budget.

With resolvable finite Monte Carlo levels, a fixed number M of Gaussian and chi-square reference draws has high-probability bounds on \(c_{F}\) and \(c_{g}\)-1. Combined with sample covariance concentration, \(n\ge C s^{-2}\log(M/\beta)\) suffices for fixed testing errors, compact parameters and a sufficient subdivision budget. Constants may depend on M and the declared component levels. The Gaussian KL lower bound still gives the matching fixed-dimensional inverse-square order.

\subsection{Calibration grid and query universe}\label{calibration-grid-and-query-universe}

The kth order reference statistic, with \(k=\lceil(M+1)(1-\alpha_r)\rceil\), is finite only when \(\alpha_r\ge1/(M+1)\). An unresolved level uses an infinite cutoff or an explicit unresolved state. Type-I control retains its event-union argument at every level. The power statement assumes resolution for every required component test, including ancestry qualification.

For a predeclared or simultaneously covered query universe Q, the graph-family error bound is

\[
\Pr\{\text{at least one false report}\}
\le\alpha_A+\sum_{q\in Q}
(\alpha_{F,q}+\alpha_{u,q}+\alpha_{x,q}).
\]

The single shared \(\alpha_A\) covers the complete qualification event. Complete coefficient-domain inversion retains the actual null coefficient pair, so its search points use the same test events. A growing query universe requires M large enough for the smallest allocated component level. In the present allocation, M=1023 resolves the variance level \(0.01/|Q|\) only for \(|Q|\le10\).

\section{Source-specific fixed-rule guarantees}\label{source-specific-fixed-rule-guarantees}

Assume each statistic, feature map and tuning rule is prespecified or selected using information invariant under the permitted null reassignment. The rank statements below apply to that rule. Selection using observed associations or working-reference scores requires calibration incorporating that selection to obtain a post-selection guarantee. The conditional Gaussian statement additionally assumes its response model and feature domain are specified for the inferential experiment.

\subsection{Randomization ranks with a fixed nuisance transform}\label{randomization-ranks-with-a-fixed-nuisance-transform}

Let \(\mathcal Z=(Z_0,\ldots,Z_{N-1})\) be an exchangeable observed-plus-reference assignment frame under a sharp absence null. Let V contain the indexed nuisance variables and responses preserved under that null. A deterministic transform of V and any statistic \(T(Z_b,V)\) preserve exchangeability of the N scores. The inclusive upper rank

\[
p=\frac{1}{N}\sum_{b=0}^{N-1}\mathbf1\{T(Z_b,V)\ge T(Z_0,V)\}
\]

is superuniform on its finite grid. Ties increase this rank. Applying the same row permutation to an exchangeable score matrix preserves exchangeability of every row-equivariant combination.

In standard light, each acquisition score is first replaced by its within-frame inclusive rank. For each reference row, multiply its eleven acquisition ranks and record whether its minimum rank meets the original acquisition-union threshold. Sorting the union-rejection part first and then sorting by rank product within each part defines one row-equivariant ordering. The union-rejection part has at most \(\lfloor Nq\rfloor\) rows, so its inclusive combined rank is at most Nq. The final combined test retains that original rejection set and has the same superuniform rank guarantee. The nuisance regression and response features in Appendix D stay fixed over this assignment frame.

\subsection{Outcome-only selection under a binary assignment law}\label{outcome-only-selection-under-a-binary-assignment-law}

Condition on the outcomes, their positions and the observed assignment count. Under the sharp absence null, all assignments allowed by the source design have the declared conditional law. The pressure selector in Appendix D depends on outcomes and positions. Its selected ordering is therefore fixed in this conditional experiment. An exact KS upper tail for that ordering is a valid original P value. Summing the 308 test allocations gives the stated 0.05 family bound. Reporting a coordinate pair by a union over its acquisitions uses this same family.

\subsection{Source-trajectory reference frames}\label{source-trajectory-reference-frames}

For the camera, the randomization unit is the target trajectory generated by its source recursion and command map. Under root absence and the qualified source law, the observed trajectory and independent source redraws are exchangeable given the nuisance and response information. Each trajectory is normalized using all 30000 source positions before its 4096-position prefix is used in the statistic. This frame supplies a superuniform condition-specific rank. The union bound gives the pair value \(\min(1,3\min_c p_c)\) and, across 50 pairs, the allocation in Table D2.

\subsection{Gaussian conditional response rows and inherited graphs}\label{gaussian-conditional-response-rows-and-inherited-graphs}

In the optical conditional model, the orthonormal contrast matrix H removes the recorded nuisance span. Since \(HH^\top=I\), the transformed error remains spherical Gaussian conditional on the design. Under a zero pixel coefficient, the additional full-model sum of squares and the full-model residual sum of squares are orthogonal Gaussian projections. Their ranks give the numerator and denominator degrees in the F statistic of Appendix D. The nine original P values therefore control additions within each acquisition at the stated union-bound level.

Let E denote the true support in a declared domain. On an event where \(\widehat C\subseteq E\), the union graph \(\widehat E_0\cup\widehat C\) has the inherited graph's false positives and at least its true positives. Thus its SHD decreases and its F1 increases whenever a previously missing true edge is added. The measured scores evaluate this mechanism against each acquisition's stated reference slots.

\section{Gaussian power settings}\label{gaussian-power-settings}

The baseline perturbation H defined in Section 6.1 has maximum absolute row sum 5/128. Gaussian centered sample scatters are generated through the Bartlett representation of the Wishart law. Every procedure in a condition receives the same scatter, and different budgets use independent sources. Each budget has one independent M=1023 calibration bank, reused across its templates, gaps and replicate sources. All calibration seeds are independent of experiment seeds. Different budgets have separate experiment sources and reference banks.

Figure S1 gives separate power curves for the four templates. The shaded intervals are pointwise 95 percent Wilson intervals for the 32 replicates at each fixed template and setting, conditional on that budget's fixed independent calibration bank. The test's finite-sample error guarantee averages over calibration and experiment randomness. The table aggregates the four template counts.

\begin{center}\begin{minipage}{\linewidth}
\centering
\includegraphics[width=\linewidth]{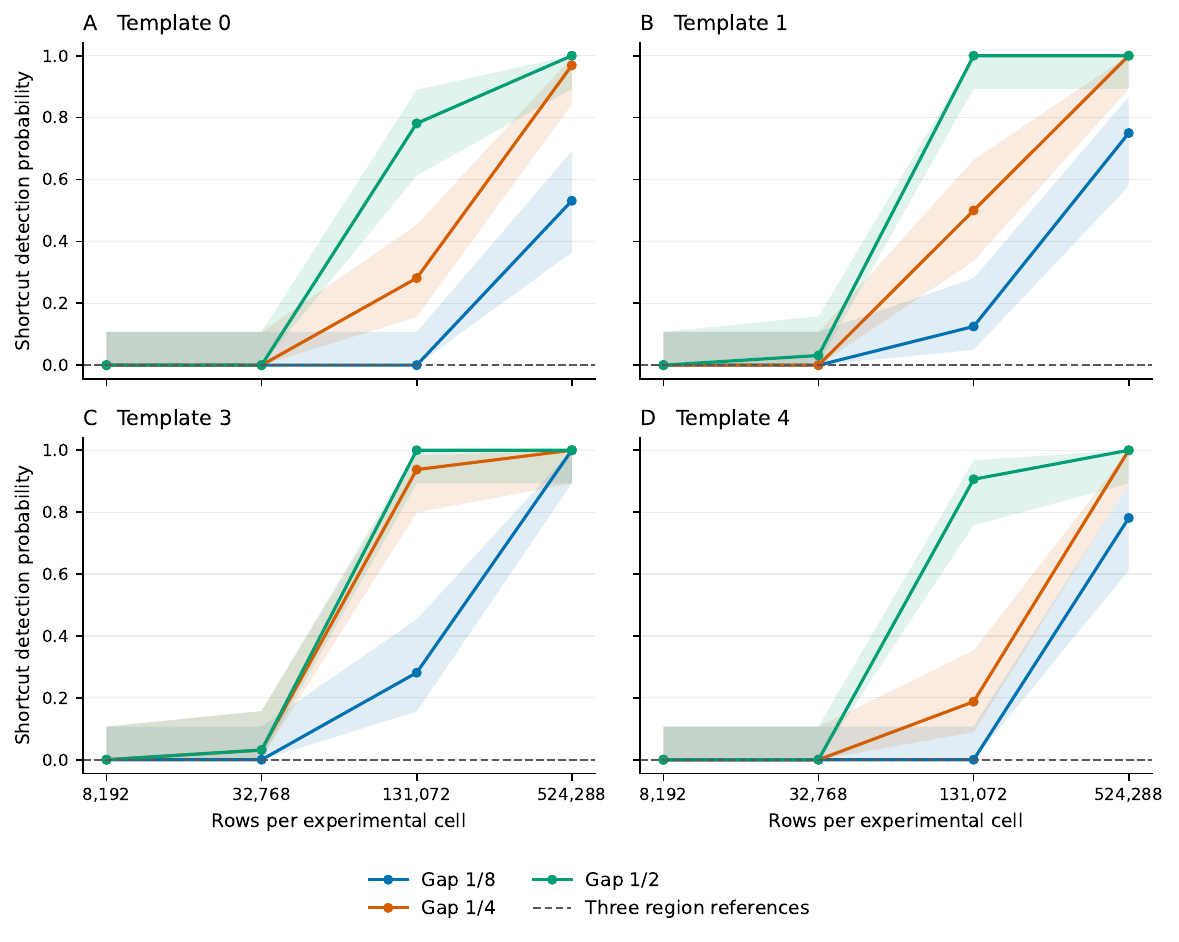}
\captionof{figure}{Shortcut detection against rows per cell for three experimental gaps. Each template-setting point uses 32 independent sources. The three reference procedures coincide at zero detections. The plots show the Gaussian covariance experiment described above.}
\label{fig:paperS1}
\end{minipage}\end{center}

Each row gives 32 independently generated conditions for one fixed template, gap and budget. Counts retain every qualification and comparator outcome.

\begingroup
\small
\begin{longtable}{@{}p{\dimexpr0.100000\linewidth-2\tabcolsep\relax}p{\dimexpr0.060000\linewidth-2\tabcolsep\relax}p{\dimexpr0.140000\linewidth-2\tabcolsep\relax}p{\dimexpr0.120000\linewidth-2\tabcolsep\relax}p{\dimexpr0.120000\linewidth-2\tabcolsep\relax}p{\dimexpr0.120000\linewidth-2\tabcolsep\relax}p{\dimexpr0.170000\linewidth-2\tabcolsep\relax}p{\dimexpr0.170000\linewidth-2\tabcolsep\relax}@{}}
\toprule
\textbf{Template} & \textbf{Gap} & \textbf{Rows per cell} & \textbf{Qualified} & \textbf{Reports} & \textbf{Original regions} & \textbf{Ordered regions} & \textbf{One Fieller box} \\
\midrule
\endfirsthead
\toprule
\textbf{Template} & \textbf{Gap} & \textbf{Rows per cell} & \textbf{Qualified} & \textbf{Reports} & \textbf{Original regions} & \textbf{Ordered regions} & \textbf{One Fieller box} \\
\midrule
\endhead
\midrule
\multicolumn{8}{r}{Continued on next page}\\
\endfoot
\bottomrule
\endlastfoot

0 & 0 & 8192 & 27 & 0 & 0 & 0 & 0 \\
0 & 0 & 32768 & 32 & 0 & 0 & 0 & 0 \\
0 & 0 & 131072 & 32 & 0 & 0 & 0 & 0 \\
0 & 0 & 524288 & 32 & 0 & 0 & 0 & 0 \\
0 & 1/8 & 8192 & 31 & 0 & 0 & 0 & 0 \\
0 & 1/8 & 32768 & 32 & 0 & 0 & 0 & 0 \\
0 & 1/8 & 131072 & 32 & 0 & 0 & 0 & 0 \\
0 & 1/8 & 524288 & 32 & 17 & 0 & 0 & 0 \\
0 & 1/4 & 8192 & 28 & 0 & 0 & 0 & 0 \\
0 & 1/4 & 32768 & 32 & 0 & 0 & 0 & 0 \\
0 & 1/4 & 131072 & 32 & 9 & 0 & 0 & 0 \\
0 & 1/4 & 524288 & 32 & 31 & 0 & 0 & 0 \\
0 & 1/2 & 8192 & 28 & 0 & 0 & 0 & 0 \\
0 & 1/2 & 32768 & 32 & 0 & 0 & 0 & 0 \\
0 & 1/2 & 131072 & 32 & 25 & 0 & 0 & 0 \\
0 & 1/2 & 524288 & 32 & 32 & 0 & 0 & 0 \\
1 & 0 & 8192 & 31 & 0 & 0 & 0 & 0 \\
1 & 0 & 32768 & 32 & 0 & 0 & 0 & 0 \\
1 & 0 & 131072 & 32 & 0 & 0 & 0 & 0 \\
1 & 0 & 524288 & 32 & 0 & 0 & 0 & 0 \\
1 & 1/8 & 8192 & 32 & 0 & 0 & 0 & 0 \\
1 & 1/8 & 32768 & 32 & 0 & 0 & 0 & 0 \\
1 & 1/8 & 131072 & 32 & 4 & 0 & 0 & 0 \\
1 & 1/8 & 524288 & 32 & 24 & 0 & 0 & 0 \\
1 & 1/4 & 8192 & 32 & 0 & 0 & 0 & 0 \\
1 & 1/4 & 32768 & 32 & 0 & 0 & 0 & 0 \\
1 & 1/4 & 131072 & 32 & 16 & 0 & 0 & 0 \\
1 & 1/4 & 524288 & 32 & 32 & 0 & 0 & 0 \\
1 & 1/2 & 8192 & 32 & 0 & 0 & 0 & 0 \\
1 & 1/2 & 32768 & 32 & 1 & 0 & 0 & 0 \\
1 & 1/2 & 131072 & 32 & 32 & 0 & 0 & 0 \\
1 & 1/2 & 524288 & 32 & 32 & 0 & 0 & 0 \\
3 & 0 & 8192 & 31 & 0 & 0 & 0 & 0 \\
3 & 0 & 32768 & 32 & 0 & 0 & 0 & 0 \\
3 & 0 & 131072 & 32 & 0 & 0 & 0 & 0 \\
3 & 0 & 524288 & 32 & 0 & 0 & 0 & 0 \\
3 & 1/8 & 8192 & 30 & 0 & 0 & 0 & 0 \\
3 & 1/8 & 32768 & 32 & 0 & 0 & 0 & 0 \\
3 & 1/8 & 131072 & 32 & 9 & 0 & 0 & 0 \\
3 & 1/8 & 524288 & 32 & 32 & 0 & 0 & 0 \\
3 & 1/4 & 8192 & 30 & 0 & 0 & 0 & 0 \\
3 & 1/4 & 32768 & 32 & 1 & 0 & 0 & 0 \\
3 & 1/4 & 131072 & 32 & 30 & 0 & 0 & 0 \\
3 & 1/4 & 524288 & 32 & 32 & 0 & 0 & 0 \\
3 & 1/2 & 8192 & 26 & 0 & 0 & 0 & 0 \\
3 & 1/2 & 32768 & 32 & 1 & 0 & 0 & 0 \\
3 & 1/2 & 131072 & 32 & 32 & 0 & 0 & 0 \\
3 & 1/2 & 524288 & 32 & 32 & 0 & 0 & 0 \\
4 & 0 & 8192 & 32 & 0 & 0 & 0 & 0 \\
4 & 0 & 32768 & 32 & 0 & 0 & 0 & 0 \\
4 & 0 & 131072 & 32 & 0 & 0 & 0 & 0 \\
4 & 0 & 524288 & 32 & 0 & 0 & 0 & 0 \\
4 & 1/8 & 8192 & 32 & 0 & 0 & 0 & 0 \\
4 & 1/8 & 32768 & 32 & 0 & 0 & 0 & 0 \\
4 & 1/8 & 131072 & 32 & 0 & 0 & 0 & 0 \\
4 & 1/8 & 524288 & 32 & 25 & 0 & 0 & 0 \\
4 & 1/4 & 8192 & 32 & 0 & 0 & 0 & 0 \\
4 & 1/4 & 32768 & 32 & 0 & 0 & 0 & 0 \\
4 & 1/4 & 131072 & 32 & 6 & 0 & 0 & 0 \\
4 & 1/4 & 524288 & 32 & 32 & 0 & 0 & 0 \\
4 & 1/2 & 8192 & 32 & 0 & 0 & 0 & 0 \\
4 & 1/2 & 32768 & 32 & 0 & 0 & 0 & 0 \\
4 & 1/2 & 131072 & 32 & 29 & 0 & 0 & 0 \\
4 & 1/2 & 524288 & 32 & 32 & 0 & 0 & 0 \\

\end{longtable}
\endgroup

\section{Joint polynomial reference and certificate checks}\label{joint-polynomial-reference-and-certificate-checks}

The joint reference uses the scalar procedure's outward Fieller components, the Fieller quadratic inequality and the four variance constraints. They form a rational nonlinear real arithmetic system on the coefficient plane. A real-arithmetic solver supplies a candidate decision, which is checked by the rational witness or exclusion calculation below. \citep{ref25}

For each of the four variance comparisons, let C=\(S_{g}\)-\(cS_{0}\) or C=\(S_{0}\)-\(cS_{g}\). The exact residual comparison has polynomial

\[
P(t,b)=C_{XX}t^2+2C_{XV}tb-2C_{XY}t
+C_{VV}b^2-2C_{VY}b+C_{YY}.
\]

Every coefficient has a rational interval obtained directly from the saved symmetric sample-covariance enclosure. In a fixed sign orthant, use lower endpoints for the constant, \(t^{2}\) and \(b^{2}\) terms. Use the lower endpoint of the tb coefficient on t\(b\ge0\) and its upper endpoint on t\(b\le0\). The linear coefficients follow the signs of t and b. The resulting lower polynomial is below every enclosed actual residual comparison throughout that orthant.

The Fieller coefficient intervals \(F_{0}\),\(F_{1}\),\(F_{2}\) give lower polynomial \(F_{0,\rm lower}+F_{1,\rm lower}b+F_{2,\rm lower}b^2\) on \(b\ge0\). On \(b\le0\) it uses \(F_{1,\rm upper}\). Append this fifth inequality and the identical outward Fieller-domain components to every orthant. The union of the four closed orthants covers the complete coefficient plane. A true missing-row pair accepted by its component tests satisfies this outer joint system. Its emptiness therefore supplies a positive certificate on the same confidence events.

The exact nonlinear real arithmetic solver returns SAT, UNSAT or UNKNOWN. A SAT outcome is checked by a rational witness against all five inequalities and the Fieller components. For an UNSAT outcome, each bounded Fieller component is covered by rational subintervals. The independent variance lower polynomials on each subinterval lie below the corresponding joint pointwise lower polynomials. Exact root isolation and sign evaluation over the full t line then prove every subinterval excluded. This verifies the entire component. A proof budget retains unfinished intervals.

The joint reference uses fixed decision and exclusion-partition budgets. Unfinished cases retain an unresolved status.

Separate rational and symbolic routines reconstruct the generated scatters, check the Fieller confidence domains and accepted joint coefficients, and establish exclusions from rational polynomial signs. They share primitive inputs and interval conventions with the scalar computation.

\section{Compatibility and core support proofs}\label{compatibility-and-core-support-proofs}

\subsection{Acyclic response compatibility}\label{acyclic-response-compatibility}

A common topological order makes A unit triangular. Then \(B=I-A^{-1}\) is strictly triangular, and congruence sends every contrast to its legal target row and column. The transformed control covariance remains positive definite. Conversely, any acyclic B yields response columns respecting its topological order, including when path contributions cancel.

\subsection{Complete core enumeration and feedback representation}\label{complete-core-enumeration-and-feedback-representation}

\textbf{Proof.} The factors are distinct, so each targeted response column has exactly the two choices in (7). Both u choices force u to precede x. Both x choices force x to precede v. Both v choices force v to precede w. The total order is therefore \(u<x<v<w\). The alternate v column also points to x, creating a cycle, so v must be physical. The order forces the unknown w column to \(e_{w}\).

If x is alternate, inversion gives \(B_{wv}\)=b and \(B_{wx}\)=(e-b)k. The u coefficient is s+(b-e)ak under physical u and (e-b)a(c-k) under alternate u. All three incoming w coefficients are nonzero. Its degree restriction therefore forces x to be physical. The two remaining u choices give precisely (9). Both matrices are compatible and acyclic. This proves the complete enumeration and the support claim.

For the feedback representation, let A' contain the alternate u,x,v columns and \(e_{w}\). Define \(W_c=\operatorname{diag}(1,1-kl,1-kl,1)(A')^{-1}\). The matrix \(B_{c}\)=I-\(W_{c}\) has coefficients

\[
\begin{aligned}
(B_c)_{xu}&=a(1-lc),&(B_c)_{xv}&=l,\\
(B_c)_{vu}&=a(c-k),&(B_c)_{vx}&=k,\\
(B_c)_{wv}&=e,&(B_c)_{wu}&=0.
\end{aligned} \tag{15}
\]

Its diagonal vanishes and its characteristic polynomial is \(z^2(z^2-kl)\). Thus its spectral radius is \(\sqrt{kl}<1\). Multiplication sends each selected alternate factor to a scalar multiple of its target basis vector, preserving (1). Finally, r=\(e_{w}\)-e \(e_{v}\) is orthogonal to every alternate factor. Hence \(r^\top(\Sigma_j-\Sigma_0)r=0\). Both x and v contrast factors have zero u coordinate, so the joint covariance of (U,\(r^\top X\)) is also unchanged across control and those two targets. Gaussianity gives the stated marginal and joint-law equalities. This completes the proof.

\subsection{Queried-edge and baseline corollaries}\label{queried-edge-and-baseline-corollaries}

\textbf{Proof of Corollary 1.} Every alternate-v choice has an x-v response cycle. With physical v, the two x choices and two u choices give the four coefficients stated in Section 4. Their signs follow from s\textgreater0, e\textless0, b\textgreater0 and k\textgreater c.~Thus each compatible DAG contains the shortcut.

\textbf{Corollary 2. Positive-definite baseline variation.} The factor and support conclusions hold for any positive-definite baseline covariance \(\Sigma_0\) for which all \(\Sigma_0+q_jS_j\) are positive definite. A sufficient condition is \(q_j\lVert S_j\rVert_{\rm op}<\lambda_{\min}(\Sigma_0)\). In particular, replacing I by I+H with symmetric \(\lVert H\rVert_{\rm op}\le1/8\) preserves the displayed normalization and gives covariance spectra in {[}1/8,15/8{]}. Contrast factorizations, stable-feedback compatibility and the residual-law equalities depend on the unchanged contrasts.

For the query u-to-w, all remaining possible parents are x and v. Including both nuisance coefficients, with zeros allowed, covers every missing-u parent set. Its four spanning-column branches are finite. Alternate-v branches contain the x-v response cycle. Physical-v branches have a nonzero u-row discrepancy under both u candidates whenever s\textgreater0.

\subsection{Uniform candidate and parent-inverse margins}\label{uniform-candidate-and-parent-inverse-margins}

The separated factors and parent inverses are uniformly Lipschitz on the compact class. Finitely many mandatory support and false-row margins have positive lower bounds. The physical-u discrepancy in the physical-x branch is s. These facts give the sufficient precision condition. Only \(a_{u}\) varies with s, and the normalization in (8) is uniformly Lipschitz with a positive denominator. The bounded covariance spectra then give the KL bound. The null physical chain at zero and the fixed strong DAG at positive s supply the testing reduction. Statistical concentration and the retained arithmetic error are separate conditions. The numerical precision is chosen so that the latter stays below c s as sample size grows.

\subsection{Branch-preserving ancestry contraction}\label{branch-preserving-ancestry-contraction}

For each target, intersect the mandatory supports of its surviving boxes. Every compatible DAG contains these response ancestor relations. Take their transitive closure. If i is a forced ancestor of j, acyclicity gives \(A_{ij}\)=0. Intersect coordinate i of every column-j box with zero and remove an incompatible box. Repeat until stable. This contraction also applies inside a selected spanning-target branch.

Each contraction is a necessary consequence of the retained regions and acyclicity. Every actual compatible DAG remains represented. In the four-node family it removes backwards response uncertainty and simplifies the x,v parent block to its triangular form.

For a selected parent branch, enclose \(\beta\) from (5). A coefficient inverse with uncertain singularity remains unresolved. Remove a candidate column only when its interval row equation excludes zero. A mandatory response edge is present when its coordinate interval excludes zero. Run Horn propagation on these mandatory supports. Report the query edge only when every missing branch is excluded.

\subsection{Root-target compatibility and measured design}\label{root-target-compatibility-and-measured-design}

Let R be known structural roots, N the remaining coordinates and \(J\subseteq R\) the known target set. For each j, intersect its candidates across all available contrasts and impose \((a_j)_R=e_j\). Assume these sets are nonempty. Write \(C=A_{NR}\) for the root-to-N response block and complete its unobserved columns freely. The full root-response block is \(A_{:R}=\binom{I_R}{C}\). Coordinates in N may themselves have empty parent sets. The model permits arbitrary real coefficients and positive-definite correlated error covariances, with every acyclic subgraph on N allowed.

For every acyclic zero-diagonal coefficient matrix T on N, put

\[
B(T)=\begin{pmatrix}0&0\\(I-T)C&T\end{pmatrix},\qquad
A(T)=\begin{pmatrix}I&0\\C&(I-T)^{-1}\end{pmatrix}.
\]

Direct multiplication gives \((I-B)A=I\). A topological order of T, preceded by R, orders B. Every root response is C, independently of T. Conversely, the root-column equation \((I-B)A_{:R}=I_{:R}\) forces \(B_{NR}=(I-B_{NN})C\), so this parameterization covers every acyclic representation with the fixed root responses.

For \(u\in J\) and \(w\in N\), its direct coefficient is

\[
B_{wu}=C_{wu}-\sum_{m\in N\setminus\{w\}}T_{wm}C_{mu}.
\]

If \(C_{wu}=0\), take T=0. If some \(C_{mu}\ne0\), take only \(T_{wm}=C_{wu}/C_{mu}\) nonzero. This is acyclic and has \(B_{wu}=0\). If all other \(C_{mu}=0\), the identity forces \(B_{wu}=C_{wu}\). This proves Proposition 2 for fixed columns and, by its independent candidate choices, for their intersection support. If \(u\in R\setminus J\), the free completion \(C_{:u}=0\) leaves the full response column \(A_{:u}=e_u\) and gives a missing-edge representation.

For every constructed W, set \(\Omega_g=W\Sigma_gW^\top\) using the original positive-definite observed covariances. These matrices remain positive definite. Each selected target factorization has \(S_j=a_jh_j^\top+h_ja_j^\top\) and \(Wa_j=e_j\). Hence

\[
WS_jW^\top=e_j(Wh_j)^\top+(Wh_j)e_j^\top,
\]

which preserves (1) and all original covariance experiments. This response-preserving construction relates to the coefficient and noise-covariance ambiguity studied in LLC. Proposition 2 supplies the explicit acyclic root-target criterion. \citep{ref22}

The zero-response condition is a population equality. When \(|N|\ge2\), an isolated response with \(C_{wu}\ne0\) can be approached by missing-edge experiments. Choose \(m\ne w\) and replace its zero u response by \(\delta\). In each contrast targeting u, replace \(a_u\) by \(a_u+\delta e_m\), keeping its partner h fixed. Every observed covariance converges to its original value and remains positive definite for sufficiently small \(\delta\). The construction \(T_{wm}=C_{wu}/\delta\) gives an acyclic missing-edge model for every nonzero \(\delta\). Thus the missing-edge class has zero covariance separation at this isolated-response experiment when coefficients and error scales are unrestricted. A finite-sample separation guarantee needs further restrictions. Its population zero pattern alone supplies no positive gap.

The A1 plan has eight known target coordinates among ten roots. Its nine readouts have no own-target acquisition. With all 19 coordinates retained, a missing row has 17 possible parents, while the eight available target columns have rank at most eight. The direct spanning-target shortcut therefore requires further information in this full-parent class. Proposition 2 explains the remaining mediated ambiguity even when all root responses are given. The measured procedures and their reference semantics are specified in Appendix D.

An exact rational check uses 24 systems of sizes four, five and six. Each checks targeted- and untargeted-root queries, target-factor recovery, acyclic zero-diagonal representations, all root responses and transformed target blocks. The existing affine-flat solver agrees on all 48 queries. The same systems verify the general T parameterization. The reproduction code supplies every matrix and decision.

\clearpage

\section{Formula-level related-work comparison}\label{formula-level-related-work-comparison}

Invariant causal prediction uses invariant residual laws and predictor--error independence to identify compatible predictor supports. Wasserstein variance minimization tests the population objective

\[
\Gamma_{-u}=\inf_{f\in\mathcal F_{-u}}\inf_\nu
\sum_g\lambda_g W_2^2\{\mathcal L(Y_g-f(X_g)),\nu\}.
\]

A common residual law gives zero objective. Martinet et al.'s Theorem 1 provides a finite-sample uniform deviation bound under its function-class and tail assumptions. Our Gaussian confidence inversion uses exact point-null pivots, complete coefficient domains and outward numerical bounds. The identifying family's marginal residual witness gives this WVM objective value zero. ICP additionally imposes its predictor--error independence conditions. \citep{ref1,ref2}

Causal Dantzig imposes inner-product invariance, expressed as equality of \(\mathbb E[X_k(Y-X^\top\beta)]\) across environments at the causal coefficient under its additive-intervention conditions. Equation (1) allows target-specific changes in error correlations. A certified ancestor order selects the observed cross-moments whose invariance follows from the unchanged off-target error block. Section 5.3 derives this transfer for a nonroot observed ancestor. \citep{ref3}

MuDo-CoM has domain covariance model \(\Sigma_g=M D_g G D_g M^\top\), where G is common and \(D_{g}\) is diagonal. Its objective matches all domain covariance matrices to this scaling model. Our target contrasts are characterized by the row-and-column restriction (1), and the identified object is the support shared by its compatible acyclic structural representations. The finite alternatives in (3) and their common support constraint specify this model's exact decision problem. \citep{ref4}

LinADMG uses an identifiable noise representation, followed by acyclicity-based resolution of permutation and scaling, to identify the structural coefficients. Its Lemma 4.1 recovers B itself under its representation assumptions. Our target-local covariance class permits two normalized response matrices with the same identified DAG support and a non-monomial ambiguity. Write \(A_{s}\) for the physical response matrix in Theorem 2 and \(A_-\) for its alternative obtained by replacing the first column \(a_{u}\) with \(v_{u}\). They satisfy

\[
A_s^{-1}A_-=I+\{(e-b)ac-s\}e_we_u^\top.
\]

The nonzero shear preserves the identified support while leaving both the normalized response and its structural coefficients ambiguous in these targeted covariance experiments. \citep{ref21}

Experimental-effect reconstruction in LLC and intervention-response inversion in inverse sparse regression provide the response-inversion lineage. We use \(B=I-A^{-1}\) inside a decision over compatible response alternatives, with support as the inferential object. \citep{ref22,ref23}

Changes-based discovery and known-target soft-intervention methods study experimental information at graph and equivalence-class level. Recent completion and orientation results use intervention-indicator semantics in maximal ancestral graphs. The present decision uses normalized linear response factors and structural row equations in the covariance model (1). Classical Horn propagation supplies the compatibility subroutine, while chi-square concentration supplies a simultaneous sampling event. \citep{ref5,ref6,ref7,ref8,ref9}